\documentclass[11pt]{article}

\newcommand{\dtcolornote}[3][]{}

\usepackage{natbib}
\let\cite\citeyear

\usepackage{setspace}
\usepackage[table]{xcolor}
\usepackage[top=1.25in, bottom=1.25in, left=1.25in, right=1.25in]{geometry}
\definecolor{darkgreen}{rgb}{0,0.6,0}
\definecolor{blue}{rgb}{0,0,0.7}%

\newcommand{\nn}[1]{\dtcolornote[Neha]{blue}{#1}}

\newcommand{\stc}{stablecoin\xspace}

\newcommand{\pv}{par-value\xspace}
\newcommand{\bd}{bank deposit\xspace}

\newcommand{\p}{par-value exchange\xspace}

\newcommand{\si}{stablecoin issuer\xspace}

\newcommand{\SI}{Stablecoin issuer\xspace}

\usepackage{graphicx}

\usepackage{amsmath}

\usepackage[english]{babel}
\usepackage{amsthm}
\usepackage{xspace}
\usepackage{subcaption}
\usepackage[capitalize,nameinlink]{cleveref}
\usepackage{booktabs, caption, makecell}
\usepackage{array}      
\usepackage{ragged2e} 

\usepackage{float}
\usepackage{parskip}
\usepackage{adjustbox}
\usepackage{tabularx}
\usepackage[T1]{fontenc}
\usepackage[utf8]{inputenc}
\usepackage{changepage}
\usepackage{titlesec}
\usepackage{tikz}
\usepackage[most]{tcolorbox}
\theoremstyle{definition}

\usepackage{epigraph}
\usepackage{etoolbox}
\usepackage{booktabs}
\usepackage{xurl}
\usepackage{multirow}
\usepackage{threeparttable}
\usepackage[table]{xcolor}
\usepackage[strict=true,debug=true]{csquotes}
\MakeOuterQuote{"}

\newtcolorbox{softframe}[1][]{
  colback=gray!10,          
  colframe=blue!50!black,     
  boxrule=0.8pt,              
  arc=3mm,                    
  left=4pt, right=4pt,        
  top=4pt, bottom=4pt,        
  #1                          
}

\usepackage{marvosym,graphicx,
xcolor,endnotes,amssymb,amsmath,amsthm,setspace}
\definecolor{rune}{HTML}{4A6672}

\usepackage{tabularx}
\usepackage{array}
\usepackage{enumitem}

\usepackage{booktabs}
\usepackage{graphicx} 

\title{The Hidden Plumbing of Stablecoins: Financial and Technological Risks in the GENIUS Act Era}

\author{Daniel J. Aronoff\thanks{MIT} \and   F. Christopher Calabia\thanks{Bank of England. Contributions reflect work completed largely while at MIT, prior to the Bank of England. The opinions expressed in the paper are the authors' own and do not represent, and should not be reported as representing, the views of the Bank of England or any of its policy committees.
} \and Anders Brownworth\footnotemark[1] \and Ashwanth Samuel\footnotemark[1] \and Neha Narula\footnotemark[1]}
\date{}

\begin{document}

\maketitle




\begin{abstract}
U.S.\ dollar stablecoins
are increasingly used as payment and settlement instruments beyond cryptocurrency markets. With the enactment of the GENIUS Act in 2025, the United States established the first comprehensive federal framework governing their issuance, backing, and supervision. This paper evaluates the financial, technological, and regulatory risks that may arise as GENIUS-compliant stablecoins scale into mainstream use. We show that maintaining par-value redemption may depend not only on backing-asset quality, but also on the functioning of Treasury and repo markets, the balance-sheet capacity of broker-dealers, and the operational reliability of blockchain-based transaction rails. Even conservatively backed stablecoins can face stress from redemption surges, market-intermediation bottlenecks, or technological disruptions. We argue that durable stability will likely require an integrated approach spanning financial-market infrastructure, prudential regulation, and software governance. While grounded in U.S.\ law, the analysis identifies principles that are relevant for regulators in other jurisdictions developing stablecoin regimes.
\end{abstract}


\onehalfspacing


\section{Introduction}

U.S.\ dollar stablecoins---digital tokens designed to maintain parity with the U.S.\ dollar and to circulate on blockchains---have rapidly grown in scale and policy relevance. 
They are widely used as settlement instruments across cryptocurrency markets and increasingly appear in emerging payment applications outside of the crypto ecosystem. 
This expansion, and forecasts of future potential, have prompted regulators to consider how instruments that resemble money in function, but not in institutional form, should be governed.

With the passage of the GENIUS Act in 2025, the United States established the first comprehensive federal framework governing the issuance, backing, and supervision of dollar-denominated stablecoins. 
The Act focuses on strengthening the quality of \si backing assets, improving transparency, and clarifying supervisory authority, with the goal of enabling the benefits of programmable, dollar-linked payment instruments while limiting obvious financial and operational risks.
%
%
While the Act substantially improves reserve quality and disclosure, it implicitly treats stablecoin stability as a balance-sheet problem, resolvable through conservative asset holdings and supervision. But a stablecoin's ability to trade at par under stress depends not only on its asset quality, but on the functioning of redemption mechanisms, markets, and operational infrastructure. The GENIUS framework leaves these stress-contingent dynamics largely unspecified. This means that stablecoins deemed "safe" could still experience instability when market liquidity, dealer balance-sheet capacity, or technical systems are strained.
%

This paper addresses these gaps by analyzing the financial, technological, and regulatory risks most likely to arise under stress as stablecoins scale in volume and are integrated more deeply into the broader financial system.
Our work is intended as a policy-oriented risk analysis rather than an assessment of stablecoins’ desirability, long-run adoption prospects, or impact on global currency arrangements.
Instead, we ask more fundamental questions: 
\textit{If stablecoins become widely used as payment and settlement instruments, what might go wrong with the guarantee of par?} 
%
%
\textit{What regulatory, institutional, operational, or technical mechanisms might help prevent destabilizing outcomes?}

To answer these questions, we analyze vulnerabilities that emerge at three interconnected layers of stablecoin architecture: the financial structure of the issuer’s balance sheet and the markets that facilitate the trade of backing assets, the technology that governs stablecoin token creation and transfer, and the regulatory framework that constrains (or fails to constrain) issuer behavior. Importantly, these layers interact in ways that can amplify stress. 
We show that ensuring a stablecoin can be redeemed for a dollar, also known as \textit{par-value exchange}, depends not only on the quality of assets backing the stablecoin, but also on the functioning of Treasury and repo markets, the balance-sheet capacity of broker-dealers, and the operational reliability of the blockchains that enact token transfer.
Even when stablecoin issuers hold conservative portfolios, frictions in market intermediation or surges in redemption demand could, under stress, test the robustness of par-value exchange.
In addition, technological risks---arising from smart contract logic, blockchain consensus mechanisms, bridges, oracles, and governance design---may impair transferability or redemption in some circumstances, potentially affecting confidence, even when reserves remain intact.

Although the analysis is grounded in U.S.\ law and institutions, the mechanisms we identify are relevant for regulators outside the United States. Stablecoins circulate globally, and many of the financial and technological channels we examine--—market-intermediated liquidation, dealer balance-sheet constraints, and blockchain operational risk—--are not jurisdiction-specific. As other authorities design or refine stablecoin regimes, the interactions documented here may inform supervisory and policy choices. 

We preview three high-level conclusions. First, asset quality alone is insufficient to guarantee par-value stability if redemption depends on intermediated markets subject to capacity constraints. Second, blockchain-based rails introduce operational risks that differ in kind from those faced by traditional payment systems and can interact with and amplify financial stress. Third, the GENIUS framework provides a strong foundation but leaves unresolved policy dilemmas concerning liquidity support, capital buffers, and redemption design that will shape outcomes in stress scenarios.



\subsection{Roadmap}
\label{subsec:roadmap}

Section~\ref{sec:key_elements} describes key elements of stablecoins under GENIUS and situates stablecoins in the hierarchy of money. We clarify the distinction between solvency and liquidity and describe the mechanics of minting new stablecoins and redemption. We then turn to the question of whether stablecoins can reliably maintain par value in times of stress. In Section~\ref{sec:insolvency} we explain how to consider the insolvency risk of stablecoin issuers given interest rate movements. In Section~\ref{sec: Balance Sheet Resilience} we compare a \si{'s} balance-sheet structure with those of commercial banks and money market funds, emphasizing differences in liability redeemability, asset liquidity, balance-sheet discretion, capitalization, and liquidity-management tools. These contrasts clarify why \si{s} may face greater challenges in maintaining par despite holding high-quality assets.

In Section~\ref{sec:redemption_risk} we turn to risks arising when issuers liquidate Treasuries to process redemptions. We examine dealer-balance-sheet bottlenecks and we review past episodes of fragility in Treasury and repo markets. These factors imply price risk to \si{s} attempting to liquidate Treasuries to meet redemption runs, and could influence the broader Treasury markets.
The section culminates in the outline of a policy dilemma: granting stablecoin issuers access to the Federal Reserve’s balance sheet could alleviate liquidity bottlenecks and reduce risk, but doing so carries significant complications for monetary policy transmission and could potentially disintermediate banks.

Stablecoins run on permissionless blockchains, which are relatively novel infrastructure in the financial system and are not yet systemically important. Section~\ref{sec:tech-and-op-risks} turns to the technical and operational risks that arise from these systems; even if a \si has the financial capacity to maintain \pv, the technological infrastructure determines whether users can actually access, transfer, and redeem their coins in practice. We discuss risks arising from smart-contract bugs, key custody, bridges, oracles, underlying blockchain consensus security, quantum computers, network congestion, and governance designs. Although some risks are in the control of the \si, many are not. We present a risk framework and suggestions for best practices to mitigate risks. In Section~\ref{sec:interactions} we explore additional channels through which technical failures can propagate into financial instability.

Section~\ref{sec:safeguards_questions} evaluates the scope and limits of the GENIUS Act as a foundation for stablecoin oversight, considering both the statutory mechanisms that promote stability and, as of yet, unresolved questions. We assess how well the Act addresses liquidity risks, redemption mechanics, market-structure vulnerabilities, and the technical fragilities we identify, and we surface areas where prudential tools, supervisory guidance, or new interagency coordination may be needed. In doing so, we situate stablecoin regulation within the broader U.S. monetary and financial architecture, clarifying how GENIUS both aligns with and departs from existing frameworks governing banks, money market funds, and other issuers of demandable liabilities. Section~\ref{sec:conclusion} concludes.



\section{Key elements of stablecoins}
\label{sec:key_elements}

A \stc is a blockchain-based asset whose issuer pledges to maintain parity with a central bank-issued currency. In this section, we first outline salient provisions of the GENIUS Act that delimit the backing asset allocations of \si{s}. We next place GENIUS-compliant \stc{s} in the monetary complex and explain the difference between issuer solvency and liquidity. After that discussion, we explain the process of redeeming and minting \stc{s}---which we will draw on in our discussion of redemption risks in Section \ref{sec:redemption_risk}.

\subsection{GENIUS provisions}

The key provisions regarding \pv and backing asset requirements in the GENIUS Act are as follows.

\begin{quote}
Sec.2(22)(A)(iii) 

(I) [The \si] is obligated to convert, redeem, or
repurchase for a fixed amount of monetary value,
not including a digital asset denominated in a
fixed amount of monetary value; and

(II) represents that such issuer will maintain,
or create the reasonable expectation that it will
maintain, a stable value relative to the value of
a fixed amount of monetary value;

Sec.4(a)(1)(A) [A \si shall] maintain identifiable reserves backing the outstanding payment stablecoins of the permitted payment stablecoin issuer on an at least 1 to 1 basis, with reserves comprising...

(i) United States coins and currency...

(ii) ...demand deposits [of banks]...

(iii) Treasury bills, notes or bonds-...

(iv)[Treasury] reverse repurchase agreements...

\end{quote}

There are three key financial requirements under GENIUS: (i) A \si must redeem its \stc into bank deposits or cash at par with eligible counterparties, i.e., \$1 of \stc can be exchanged for \$1 of cash or bank deposits, (ii) A \si is required to maintain an expectation that its \stc will trade at \pv with the dollar and (iii) the \si must secure the value of its \stc liabilities with backing assets composed of cash, \bd{s}, or US Treasury liabilities (``Treasuries'').\footnote{Treasuries can include direct ownership, repos and futures.} 
We explore how the allowed scope of backing assets, (iii), affects the ability to maintain \pv, (i) and (ii). 

\textbf{Other \stc attributes} There are some types of \stc{s} that are explicitly excluded under GENIUS. For example, algorithmic \stc{s} that rely on arbitrage mechanisms between cryptocurrencies, rather than backing assets, to maintain \pv are excluded. Terra-Luna, which collapsed in 2022, is the most prominent member of this group  \citep{Liu2023Anatomy}.\footnote{GENIUS (Sec.14) mandates a study of algorithmic \stc{s} for potential future regulatory approval.}  In addition, some properties appear to be left to the discretion of \si{s}, such as the protocol chosen by the \si for redeeming \stc{s}. At one end, every \stc owner can redeem directly with the \si. At the other end, the \si designates a small number intermediaries to act as market-makers in purchasing \stc{s} and then either warehousing and re-selling at a later date, or redeeming with the \si. Currently, the predominant \si{s} (Circle and Tether) use intermediaries. 

\subsection{The position of stablecoins in the hierarchy of money}
\label{subsec:hierarchy_of_money}

To facilitate our discussion of \stc financial risks, we describe the position occupied by \stc{s} in the hierarchy of the U.S. monetary system. The hierarchy is composed of interlocking sectoral balance-sheets in which the liabilities of one sector are securities of, and settlement currency for, the adjacent sector lower down the hierarchy  \citep{Mehrling2012Hierarchy}. At the top of the hierarchy is the central bank (the Federal Reserve), with assets comprised primarily of U.S. government debt, and liabilities comprised of reserves and cash. 

Next in the hierarchy are banks, with assets comprised of central bank liabilities---reserves and cash---alongside loans. Inter-bank payments are settled by transferring reserves. Banks issue deposit liabilities, which they guarantee to convert into cash at par, i.e., \$1 of a bank deposit is exchangeable for a \$1 cash liability of the Fed. Bank deposits (and cash) are assets of, and money for, the non-financial sector, which in turn issues securities. These are liabilities whereby the issuer promises to pay holders scheduled amounts of bank deposits at future dates. What ties the hierarchy together is the ability to convert the liabilities at one level, e.g., \bd{s}, into liabilities of the next level up, e.g., cash at short notice at par (i.e., on demand). Notably, it is not the actual volume of conversions that enforce par, but the \textit{option to convert}. \footnote{``Private bank deposits remain at par with currency because of the guaranteed option of conversion into central bank money. It is the existence of the option, not the frequency of its use, that keeps the system anchored'' \citep{goodhart1988}. We can add that convertibility in the other direction, from cash into \bd{s}, is also required.}

\begin{figure}[H]
\begin{center}
\includegraphics[page=1,width=0.9\textwidth,height = 0.3\textheight]{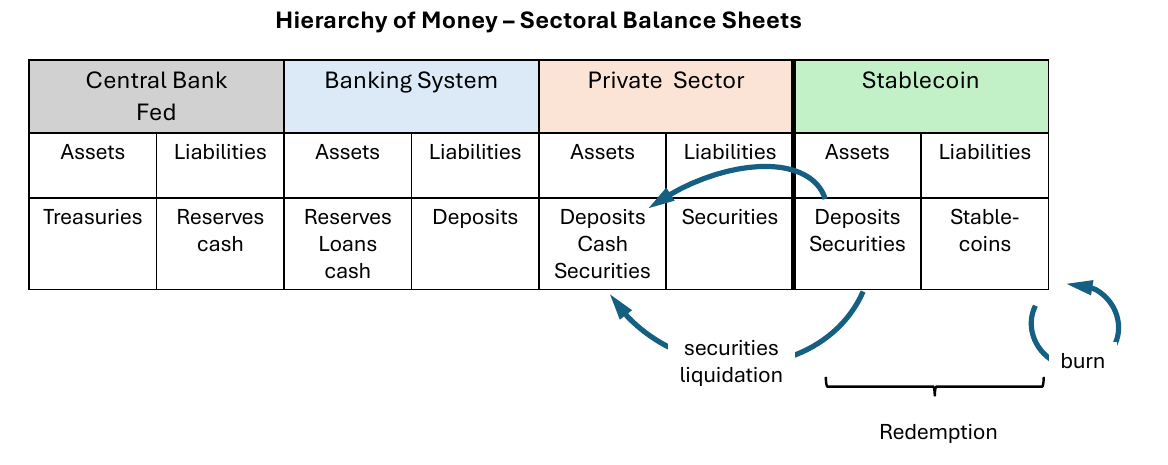}
\end{center}
\caption{A securities-backed \stc is redeemed by liquidating securities to obtain bank deposits.}
\label{fig:Hierarchy_of_Money}
\end{figure}

\textbf{Stablecoins are monetary liabilities that lie lower in the hierarchy than deposits and cash}, because redemption normally settles in commercial bank deposits and can require market intermediation to transform securities into deposits. This placement, however, is not automatic. It is a function of institutional design: if stablecoin issuers can hold central bank reserves as a backing asset and can settle redemptions and payments in central bank money without relying on sponsoring commercial banks, stablecoins can approach the same position in the hierarchy as commercial bank money. \footnote{The Bank of England  has proposed a design that allows \si{s} to hold central bank reserves and borrow from the central bank \citep{BoE2025}. In Section \ref{subsec:dilemmas} we discuss tradeoffs, in terms of cost and other tradeoffs, this arrangement implies.} By contrast, when redemption settlement relies on commercial bank deposits and dealer-intermediated liquidation of securities, as is the case under GENIUS, stablecoins remain lower in the hierarchy even when the reserve portfolio is high quality. Figure \ref{fig:Hierarchy_of_Money} displays the position of GENIUS \stc{s} in the hierarchy and outlines the redemption process.


\subsection{Liquidity and solvency}
\label{subsec:liquidity}

There are two key elements implicit in the definition of \pv exchange. One is \textit{solvency}; the market value of backing assets must exceed \stc liabilities. The other is \textit{liquidity}; the backing assets, to the extent they are composed of Treasury securities,  must be sold in a timely manner to meet redemptions.  Securities vary in terms of liquidity; a comparatively more liquid security is defined as ``being more certainly realizable [into bank deposits or cash] at short notice without loss'' \citep{keynes1930}. The most liquid securities are Treasury bills, Treasury repos, and money market fund shares. 
Solvency is a necessary, but not sufficient, condition to maintain \pv exchange: execution requires liquidity. Most of the time, redemptions are processed without delay at \pv. In a stress event where redemption demand spikes, Treasury market infrastructure---broker-dealer market-making capacity, trading software systems, etc.---can become overwhelmed by the flood of selling orders, leading to price collapse and/or delayed transaction processing.\footnote{In Section \ref{subsec:fragility} we review recent episodes of stress in the US Treasury market.} It is at those critical moments of market stress that \pv is tested. At such times, the ability to maintain \p is determined by the liquidity of the securities that the \si must sell to redeem its \stc{s}.

This motivates a conceptual partition of GENIUS-authorized \stc{s} into those backed by \bd{s} and those backed by securities. The ``convenience yield'' of bank deposits---arising from their superior payment and liquidity services---causes them to yield less than other securities \citep{mishkin, HolmstromTirole, vayanosvila}. The convenience yield for a \si arises from the \stc redemption process. The yield spread between Treasury securities and \bd{s} provides incentive for the \si to hold the former as its primary backing asset, which implies a tradeoff of yield for liquidity. The \si can immediately redeem with a \bd  it holds as a backing asset. By contrast, when the backing asset is a security, the  \si has the added step of liquidating the security into a \bd to redeem. Selling the Treasury adds a layer of risk and delay. Bank deposits can be transferred instantaneously with FedNow, whereas Treasury transactions require one day (T+1) to settle. Two things to note about this partition are the following: 

\begin{itemize}
\item A \si can operate on both sides of the partition if it holds \bd{s} and securities, as most do. When it redeems using pre-existing deposits, the \stc{s} are \bd-linked. When it liquidates a security to obtain a \bd, the \stc{s} are securities-backed. 

\item The distinction is based on liquidity, not solvency. A \bd is not solvent, and may not be honored, if its bank is insolvent. This risk was underscored in 2023 when Circle, the issuer of \stc USDC,  was temporarily unable to transfer its \$3.3 bn of uninsured deposits held at Silicon Valley Bank (SVB), which had become insolvent.\footnote{In the event, Circle's \bd account was effectively bailed out by the Fed\\
\url{https://www.financemagnates.com/cryptocurrency/svb-crisis-circle-escapes-usdc-depeg-with-regulatory-assurance/}} 
\end{itemize}

\subsection{Stablecoin redemption and minting mechanics}
\label{subsec:stablecoin_supply}

As discussed in Section \ref{subsec:hierarchy_of_money}, \pv is enforced by the option to convert the lower form of money, in this case \stc{s}, into the higher form of money, in this case \bd{s}, and back again.\footnote{We use ``minting'' to denote the entry of \stc{s} into circulation. This is a slight abuse of the term, since a \si, or its distributor, can retain minted \stc{s} as reserves to meet future demand. This has been a common practice.} Here we briefly describe the mechanics of how minting and redemption work assuming the \si is not simply using bank deposits to back its tokens.
Note that, perhaps counterintuitively, the process of redeeming and minting \stc{s} does not directly involve a change in the overall volume of \bd{s}.\footnote{We do not here explore the possibility of second-order effects on \bd{s} such as changes in monetary policy prompted by growth of \stc{s} that may cause the volume of reserves, and thereby deposits, to be altered.}

\begin{figure}[H]
\begin{center}
\includegraphics[page=1,width=0.85\textwidth,height = 0.3\textheight]{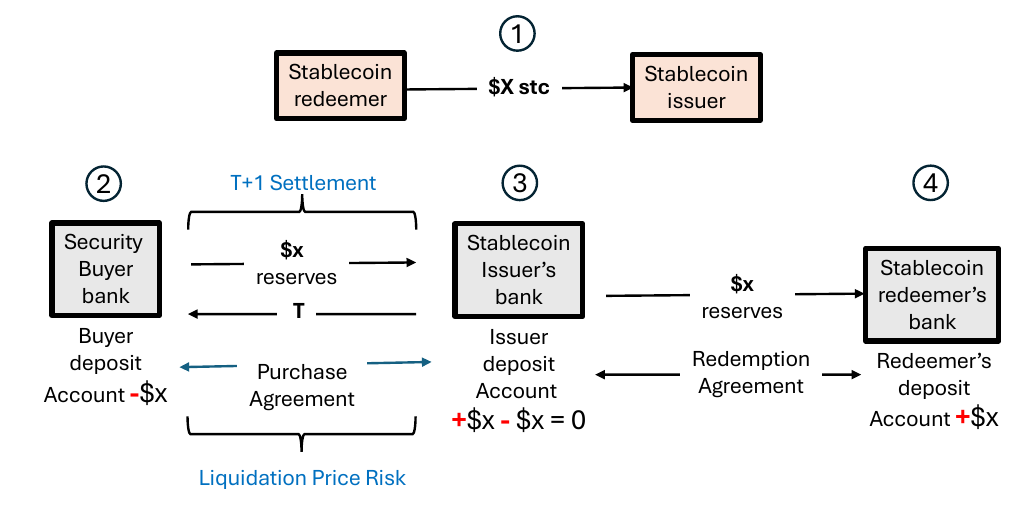}
\end{center}
\caption{Redemption \stc payments flow}
\label{fig:redemptionstc}
\end{figure}

\textbf{Redemption.} Figure \ref{fig:redemptionstc} depicts the redemption process when the \si  holds securities as the backing asset and liquidates Treasuries (denoted $T$). The sequential order is as follows.

\begin{enumerate}
\item The redeemer sends the \stc with value $\$x$ to the \si in a blockchain-based smart contract. The \si then removes it from circulation. 

\item The \si sells Treasuries. This results in the buyer of the Treasuries sending (i.e., subtracting)  $\$x$ from its \bd account in return for receipt of $T$ Treasuries. The buyer's bank transfers $\$x$ of reserves to the \si{'s} bank in exchange for the transfer of \bd liability. The buyer will typically be a broker-dealer in the secondary market for Treasuries. 

\item The \si sends $T$ to the buyer and receives an increase of \$x in its \bd account. Thereupon, it sends $\$x$ to the redeemer's \bd account, which reduces its \bd account to the initial level. The \si{'s} bank transfers \$x of reserves to the redeemer's bank in exchange for the transfer of \bd liability.

\item The redeemer receives an increase of $\$x$ to her \bd account.
\end{enumerate}

At the end of the redemption, the redeemer has exchanged a \stc for an equal value increase in her \bd account, and the  \si has reduced its liabilities and assets each by $\$x$. Note, however, an orderly redemption requires that the Treasury can be sold ``at short notice without loss'' (which still requires T+1 to settle). In Section \ref{sec:redemption_risk} we explore the viability of this assumption. Finally, note that redemption does not alter the volume of \bd{s}. The value $\$x$ moves from the deposit account of the Treasury buyer to the redeemer's account (via the \si{s} \bd account).\footnote{In the event the \si redeems with \bd{s} it holds as a backing asset, Step 2 is eliminated and the transfer moves directly from the \si{s} \bd to the redeemer's \bd.}

\begin{figure}[H]
\begin{center}
\includegraphics[page=1,width=0.65\textwidth,height = 0.25\textheight]{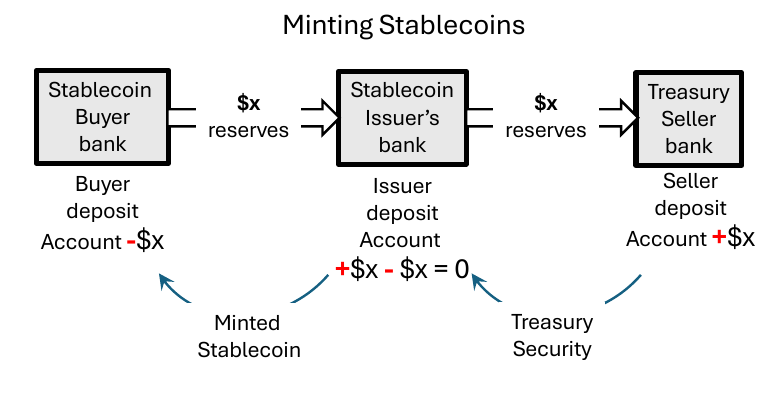}
\end{center}
\caption{Minting \stc flow}
\label{fig:minting_process}
\end{figure}

\textbf{Minting.} Figure \ref{fig:minting_process} depicts the flow of central bank reserves and deposits that takes place when a new \stc is minted. The interpretation of the chart is similar to Figure \ref{fig:redemptionstc}. Moving from left to right, the buyer sends an $\$x$ deposit to the \si for the purchase. If the \si elects to hold a Treasury security, it sends the deposit to the seller of the Treasury security as payment for the security. The deposit value moves between accounts, but the net value is unchanged. As with redemptions, minting generates no change in \bd{s}. The deposit value $\$x$ moves from the buyer's \bd account to the Treasury seller's account.  At some point during this process, the \si issues a blockchain transaction to credit the buyer's account in a smart contract on a public blockchain.

\textbf{Direct vs. indirect access to minting and redemption.} At the present time, the two dominant \si{s}, Circle and Tether, do not offer minting and redemption services to anyone. Instead, these services are limited to sets of intermediaries who in turn interact with stablecoin buyers and sellers on a secondary market (primarily cryptocurrency exchanges).\footnote{Notably, the EU MiCA \stc regulations require direct access.}  Ma et al. \cite{ma2025stablecoin} present an interesting tradeoff between par and run-risk: With direct access, there is no incentive to trade in a secondary market for \stc{s} because every owner has the option to redeem at par.  However, this increases the incentive to redeem at times of market stress, thereby increasing run risk. On the other hand, indirect access induces a secondary market with prices that can deviate from par, but reduces the incentive to run precisely because \pv is not guaranteed, even to the early movers. 

\subsection{Challenge maintaining par-value without open market operations}

As far as we know, today the largest \si{s} do not engage in open market operations to maintain the exchange rate of their stablecoin tokens on secondary markets, like cryptocurrency exchanges. However, we note that depending on how strictly the  \pv requirement of GENIUS (Sec 2(22)A(iii)) is interpreted and enforced, \si{s} might need to exercise lesser or greater control over the volume of \stc{s} in circulation.

A  \pv requirement can be analyzed along two dimensions: the goal of the requirement and the methods to achieve the goal. \textit{The goal} is to enable an owner of a \stc to redeem or sell her \stc for a fixed dollar amount.  This can be interpreted in at least two ways. One way is to require that the \stc can be sold ``on short notice without loss [i.e., at par]’’.  This is a rigorously fixed exchange rate. It can alternatively be interpreted as allowing the exchange value to fluctuate within defined bounds, above and below \pv. This is a corridor target. 

\textit{The method to achieve the goal} has to do with how the \si adjusts the supply of its \stc{s}, where the circulating supply of a \stc at a point in time is the difference between the cumulative quantity minted and redeemed by the \si. There are three ways in which a \si can intervene to alter supply:

\begin{enumerate}
\item Directly mint or redeem \stc{s} based on requests from buyers and sellers at par-value (specified by MiCA)
\item Mint or redeem from intermediaries at par-value, who, in turn, buy or sell on secondary markets (how Circle and Tether currently operate)
\item Create new stablecoins and offer them for sale without an initiating request, or purchase \stc{s} on secondary markets\footnote{There are two circumstances where minting in response to a demand for \stc{s} may not be feasible, even as the value of the \stc is above par. One is during a financial crisis where a contraction of credit limits the volume of \stc{s} that arbitrageurs are able to purchase. The other is where the interest rate on \bd{s} and Treasury securities drop below the level required for the \si to make a profit, for example if interest rates are negative. In that case, the \si will be reluctant to mint. These constitute---possibly irreducible---gaps in the ability of a \si to maintain \pv.} 
\end{enumerate}

We do not know the specific goal that will ultimately be enforced (a rigorously fixed exchange rate, a corridor target, or merely a best-effort attempt at par), or whether there will be directions on the use of methods. We can, however, comment on the difference in discretionary control that the \si will have under the three types of goals.

\textbf{Rigorous fixed exchange rate.} Under strict enforcement of \pv, the supply of \stc{s} would be determined by the demand to acquire or to redeem \stc{s}. The \si would passively adjust supply to these market forces (method 3). The reason is that, to maintain par at every moment in time, the \si would need to make a standing offer to redeem or purchase when the \stc is below target, to reduce supply to accommodate market demand to convert \stc{s} into dollars, and mint when it is above target, to accommodate market demand to convert dollars into \stc{s}. This places the \si in a position similar to a central bank that maintains a fixed exchange rate with an offshore currency. It is an axiom of monetary theory that the domestic money supply is endogenously determined by the requirement to maintain par when the central bank commits to a fixed exchange rate with another currency or asset (such as gold).\footnote{``...under fixed exchange rates the central bank does not control the money supply since it must fix the exchange rate.'' (Krugman et al. \cite{Krugman2012}).} 

\textbf{Corridor around \pv.}  A corridor, which establishes a range of exchange values, would  provide the \si discretionary control over the \stc supply inside the corridor. Here the \si can allow the  exchange rate to drift above par without minting, or allow it drop below par without redeeming. Outside the corridor market forces will dictate the adjustments to supply the \si will need to effectuate, engaging in method 3. 

\textbf{No strict enforcement of \pv.} In this case, GENIUS is interpreted to require \si{s} to provide \pv exchange only with the set of intermediaries who have access to direct minting and redemption with the \si. Many \stc holders might experience a loss of \pv. 

 


\section{Insolvency risk}
\label{sec:insolvency}

The GENIUS Act requires a \si to maintain backing assets at least equal in value to its \stc liabilities (Sec.4(a)(1)(A)).\footnote{The GENIUS Act (Section 4.(4)(A)) directs certain regulatory agencies to issue capital and liquidity requirements for \si{s}, which may, inter alia, mandate an increase in the 1:1 capital requirement. We do not here speculate on the content of future rules.} It is a knife-edge requirement. A \si that ``just'' meets the requirement at one moment will become insolvent the next moment if the market price of its backing assets falls by any amount. This is a nontrivial risk as Treasuries are subject to interest rate risk, i.e., their value moves inversely to their yields. In this section, we explain how the two dominant \si{s}, Tether and Circle, have mitigated interest rate risk by allocating the majority of their backing asset into short duration securities. This minimizes interest rate risk and maximizes flexibility to liquidate securities to meet redemption demands. Equation \ref{eq:value_difference} shows the progression of a \si'{s} backing assets value over time.\footnote{GENIUS prohibits \si{s} from paying interest on liabilities. This eliminates a source of solvency risk that is present in banks, who pay interest on deposits and other liabilities.}   

\begin{equation}
\label{eq:value_difference}
A_{t+1} = A_{t} + \underset{\text{Treasury securities}}{\underbrace{\underset{\text{interest}}{\underbrace{r_{t,T}T_{t}}} + \underset{\text{capital gain}}{\underbrace{(T_{t+1} - T_{t})}}}} + \underset{\text{\bd{s}}}{\underbrace{\underset{\text{interest}}{\underbrace{r_{t,D}D_{t}}} + \underset{\text{capital gain}}{\underbrace{(D_{t+1} - D_{t})}}}}
\end{equation}

In Equation \ref{eq:value_difference} time is partitioned into a sequence of periods denoted $t$, $t+1$ etc...$A_{i}$ is the value of the \si{'s} backing assets at the beginning of period $i$; $T_{i}$ and $D_{i}$ are the \si'{s} investment in Treasuries and \bd{s} at the beginning of period $i$ and $r_{i,T}$ and $r_{i,D}$ is the interest paid on Treasuries and \bd{s} in period $i$, which is fixed at the beginning of period $t$. The \textit{level} of interest rates earned on backing assets---so long as it exceeds the \si{'s} operating cost---adds to the value of backing assets. The primary source of insolvency risk is from a capital loss, i.e., $T_{t+1} < T_{t}$. Capital loss is caused by a decline in interest rates from the beginning of period $t$ to the beginning of period $t+1$. In other words, capital loss is determined by \textit{timepath} of interest rates on backing assets.\footnote{Here we treat backing assets under GENIUS as ``safe assets'' and ignore default risk.} GENIUS requires that a study of possible capital rules be undertaken but at present there is no mandate. In this case it is left to the \si{s} to determine their own investment and retention policies. 

\subsection{Backing asset allocations}

The reported asset allocations of the two dominant \si{s}, Circle and Tether, are concentrated in short duration securities, notably overnight reverse repos and Treasury bills.\footnote{We do not address controversies surrounding the veracity of the reported asset positions.} In terms of equation \ref{eq:value_difference}, the interval between $t$ and $t+1$ is short, which minimizes the risk of capital loss.

\begin{table}[htbp]
\centering
\begin{adjustbox}{max width=\textwidth}
\begin{tabular}{l|ccccc} 
\toprule
& \makecell{Repo} & \makecell{Treasury\\bills} & \makecell{Bank\\deposits} & \makecell{Bitcoin} & \makecell{other}\\
\midrule
Circle & 43 & 24 & 11 &&\\
\hline
Tether & 21 & 113 & 6 & 10 & 18\\
\bottomrule
\end{tabular}
\end{adjustbox}
\caption{Backing Asset Allocations Circle and Tether Q3 2025 (in billions USD).\protect\footnotemark}
\label{tab:asset-allocations}
\end{table}
\footnotetext{Figures rounded to the nearest billion USD.
Circle source: Deloitte Independent Accountant's Report, November 26, 2025.
Tether source: BDO Independent Auditors' Report, October 31, 2025. Note that Tether's figures include non-US domiciled \stc{s}, which lie outside the jurisdiction of GENIUS.}

\subsection{Reverse-repos provide a capital cushion}

Here we explain how reverse-repos provide the \si with maximal protection against capital loss. In a reverse-repo trade, the \si lends \bd money in the repo market and receives Treasuries in exchange, with an agreement to resell those Treasuries at a later date. A repo trade is composed of two contracts entered into at the same time. A contract for party A to purchase Treasuries from party B today (the "first-leg''), and a contract for party B to re-purchase the same type---CUSIP or general collateral---and volume of Treasuries from party A at a future date (the "second-leg''). Effectively, party A is making a collateralized loan to party B. GENIUS allows \si{s} to hold reverse Treasury repos.

\begin{figure}[H]
\begin{center}
\includegraphics[page=1,width=0.6\textwidth,height = 0.35\textheight]{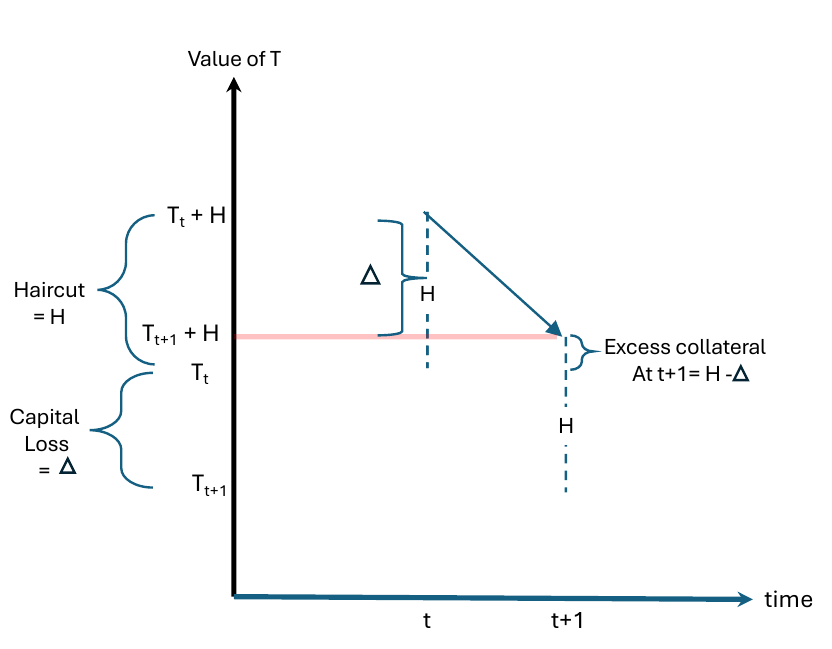}
\end{center}
\caption{Repo Lending with Haircut}
\label{fig:haircut_graph}
\end{figure}

Typically, repo borrowers---the counterparties to \si{s} in the repo trade---provide excess Treasury collateral, referred to as the ``haircut''.\footnote{For example, the standard haircut in the tri-party repo market is 2\% \citep{Paddrik2021TripartyRepo}. However, there may not be a haircut when the counterparties have another trade that offsets the risk of the repo trade \citep{ Hempel_uncleared}.}  The haircut provides the \si insurance against capital loss. The short time period of the overnight trade (24 hours), or the twice-daily re-margining that applies to longer duration repos,  limits the size of tail---i.e., extremely large---capital loss. For a haircut of size $H$---meaning that the Treasuries held by the \si exceed the value of its first-leg purchase by $H$---the Treasury can be liquidated and the \si will receive back its purchase price for any decline in market price up to $H$.  Figure \ref{fig:haircut_graph} displays the effect of a haircut in a repo lending trade (it omits interest payments for simplicity). The vertical axis is the market price of Treasuries. The horizontal axis is time $t$. 

It depicts a decline in value between the first-leg at time $t$ and the second-leg at time $t+1$, of $\Delta$ where the haircut $H$ (i.e., the value of excess collateral) exceeds the price decline ( $H > \Delta$). Without the excess collateral $H$, the value of the Treasuries would drop below the purchase price. This outcome would occur if the \si purchased Treasuries in the secondary market. Treasury repos add two layers of protection. One layer is that the repo counterparty has a contractual obligation to repurchase at the second-leg, notwithstanding a decline in market price. The second layer is that the haircut adds excess collateral, which creates a cushion that enables the \si to liquidate, if necessary, and recover its first-leg purchase price, up to a decline of $H$.

\textbf{Observations} We make three observations regarding the protection against capital loss provided by reverse repos.

\begin{enumerate}
\item The counterparty has a contractual obligation to repurchase the Treasury at the negotiated price, regardless of any decline in its market value.

\item If the \si{'s} counterparty fails to repurchase, the \si can liquidate the Treasury in the secondary market and recover its first-leg purchase price, for any decline in market value of the Treasuries collateral up to the haircut.

\item The \si can mitigate its risk of capital loss by sequencing overnight repo trades---``rolling over.'' When Treasury yields are increasing (i.e., price is declining) the \si receives back its initial purchase price at the second-leg and re-lends (i.e., enters into a new reverse repo trade) its money at a higher interest rate the next period, and so on. 
\end{enumerate}

\subsection{Summary}

GENIUS does not (yet) promulgate a capital adequacy standard for \si{s}. The two largest issuers have chosen to mitigate their insolvency risk by investing in short duration Treasury securities. Opinions may differ on whether that is an adequate cushion against capital loss. There are two paths forward concerning \si capital standards. One way is for regulators to establish capital adequacy rules that address the risk of insolvency (which we address in Section~\ref{sec: Balance Sheet Resilience}) and illiquidity (which we address in Section \ref{sec:redemption_risk}). The other way is to allow \si{s} to choose their backing asset portfolios and compete for acceptance by the public, subject to the constraints in the extant legislation.


\section{Building a resilient balance sheet}
\label{sec: Balance Sheet Resilience}

In the prior section, we explored how two stablecoin issuers sought to mitigate their exposure to interest rate risk so that their backing assets remain at least equal in value to their stablecoin liabilities and they can honor redemptions at par. Now we broaden our scope to consider how stablecoin issuers hedge against unexpected losses stemming from all risks, including operational and broader market rate risks. The design of a stablecoin issuer's balance sheet differs markedly from the most familiar financial institutions that offer similar \pv promises, namely commercial banks and money market mutual funds (MMFs). Understanding the structural differences in balance sheet designs is critical for assessing the solvency, liquidity, and run risks associated with stablecoin issuers, even for those that are aligned with the GENIUS Act.

To keep the promise of \pv redemptions, stablecoin issuers must invest proceeds from coin sales in assets that preserve value and can be liquidated quickly at low cost. Under the GENIUS Act, licensed issuers are largely confined to cash (in U.S.\ dollars), bank deposits, and short-term U.S.\ Treasury-related securities \citep{12USC5901}. These constraints are intended to limit credit risk in the reserve portfolio, but they do not eliminate exposure to interest-rate, operational, or broader market risks.\footnote{\protect\citet{liang2025stablecoins} notes that the Act does not distinguish between holdings of insured versus uninsured deposits, the latter of which may increase an issuer's exposure to run risk.}

To date, stablecoin issuers have generally not been subject to bank-like capital requirements. The GENIUS Act empowers regulators to issue capital and liquidity requirements that are "tailored to the business model and risk profile of permitted payment stablecoin issuers." Regulators have not yet issued these requirements, but the law appears to restrict U.S. regulators from applying the minimum leverage and risk-weighted capital requirements that govern commercial banks under the ``Collins Amendment'' directly to stablecoin issuers \citep{12USC5901-4c2}; see also \citep{sullivan2025stablecoin}. A central question is therefore whether a portfolio of high-quality, rapidly turning over assets provides sufficient protection against losses from interest-rate, operational, and market risks. If not, explicit capital buffers or other financial resources may be needed to ensure that stablecoin issuers remain operational and honor par redemptions. 

For stablecoin issuers, this analysis is complicated by their reliance on novel technologies. Disruptions in the underlying technology do not directly alter the market value of reserve assets such as U.S.\ Treasuries. Frictions or failures in the systems used to transfer those assets or convert them into cash could nonetheless undermine confidence in the stablecoin, depress its market price, and impair the issuer's ability to redeem at par. We discuss in detail these technological and operational risks in Section~\ref{sec:tech-and-op-risks}.

Prudent balance sheet design and management of reserve backing assets, liabilities to \stc owners, and the issuer's own funds are therefore necessary but not sufficient conditions for preserving the \pv exchange. We will draw on key risk measurement and mitigation practices that commercial banks and money market mutual funds or their regulators use and compare their balance sheets with those of stablecoin issuers to understand their respective risks.

\subsection{Comparing stablecoin issuers to commercial banks}

When a stablecoin issuer receives cash from coin buyers, it records the proceeds as liabilities and invests them in reserve assets. These assets are chosen for their perceived safety (asset quality) and their ability to be sold quickly for an amount at least equal to the requested redemption (liquidity). Under the GENIUS Act, the issuer retains the income on reserve assets and does not share it with \stc owners, much as in some traditional intermediation arrangements. This resemblance has led some commentators to view stablecoins as equivalent to bank deposits; see \citep{armstrong2025stablecoins} in the \emph{Financial Times}.

The parallel with banks is useful but incomplete. We highlight five characteristics that differentiate the liquidity and strength of stablecoin issuers' balance sheets from those of commercial banks (redeemability, liquidity of assets, control over balance sheet growth, access to public support, and capital cushions) and assess the implications for resilience and par-value redemption.

\textbf{Redeemable liabilities.} Stablecoin issuers are more exposed to bank-style runs because nearly all of their liabilities are redeemable on demand. Commercial banks, by contrast, fund themselves with a mix of demandable and non-demandable liabilities. In U.S.\ aggregate data for 2024, checking deposits (17\%) and overnight repos and federal funds (10\%) are payable on demand, while savings deposits (35\%) are not \citep{federalreserve2024z1} By this simple metric, banks are more insulated from immediate withdrawals than stablecoin issuers.

\textbf{Liquidity of assets.} Commercial bank assets are much less liquid than the assets held in stablecoin reserve portfolios. Loans account for roughly half of bank assets and are neither payable on demand nor easily sold quickly without incurring losses. Stablecoin reserves, in contrast, typically consist of U.S.\ Treasury securities, repos, and bank deposits or cash. Treasury-related instruments and cash can usually be liquidated rapidly and at or near par, provided that Treasury markets remain orderly and stablecoin issuers retain access to them. This structure suggests that stablecoin issuers may be better positioned than banks to meet sudden surges in redemption requests, although this conclusion depends critically on the behavior of Treasury markets under stress, which we examine in Section~\ref{sec:redemption_risk}.

\textbf{Control over balance sheet growth.} Commercial banks have more discretion over their balance sheet growth. They can choose whether to expand or contract loan portfolios, decide which borrowers to serve, and adjust the composition of assets and liabilities over time. Stablecoin issuers, in contrast, have limited control over the size of their balance sheets: to prevent their coins from trading above par, they must generally accommodate demand for new coins by accepting cash and expanding reserves. Their balance sheets will tend to expand when demand is high and contract when demand is low. The GENIUS Act restricts their investment choices largely to Treasury-related securities and deposits at commercial banks.

\textbf{Access to public support.} Commercial banks benefit from multiple forms of explicit and implicit public support. Eligible deposits are insured by the Federal Deposit Insurance Corporation (FDIC), for which banks pay a premium but in return obtain a strong guarantee that bolsters depositor confidence. Banks also have access to central bank liquidity through the Federal Reserve's discount window and standing repo facilities, especially in times of stress. Under the GENIUS Act, licensed stablecoin issuers do not receive deposit insurance and currently lack routine access to such public backstops, leaving them more reliant on their own assets to meet unexpected losses or rapid redemption waves.\footnote{For comparison, the Bank of England's November 2025 consultation paper \citep{boe2025stablecoins} proposes that systemic sterling-denominated stablecoin issuers hold 40\% of reserve assets in non-remunerated central bank deposits and notes that the Bank is ``considering providing access to a backstop lending facility for eligible, solvent, and viable systemic stablecoin issuers.''}

\textbf{Capital and solvency.} Banks and stablecoin issuers differ sharply in how they use capital to absorb losses. Bank capital consists of equity---funds raised through stock issuance or retained earnings---that stands junior to deposits and other liabilities. Accounting identities require assets to equal liabilities plus capital. A bank funded entirely by deposits with no capital buffer would become insolvent if any borrower unexpectedly defaulted on a loan. Sufficient capital therefore protects solvency and enhances resilience; see \citet{farag2013bank} for a primer.

Capital requirements are set in part by regulation. The GENIUS Act establishes a Stablecoin Certification Review Committee (SCRC) comprising the federal banking regulators, chaired by the Secretary of the Treasury, to develop the regulatory framework for licensed issuers. The same agencies---the Federal Reserve, the Office of the Comptroller of the Currency (OCC), and the FDIC---set capital and liquidity standards for commercial banks, drawing on international guidance from the Basel Committee on Banking Supervision. 

(Because issuers with less than \$10 billion in coins outstanding may opt for state-level regulation, the GENIUS Act requires the Secretary of the Treasury to ``establish principles'' for determining the equivalence of a state's regulatory framework with the federal approach. See 12 U.S.C.§5903(c)(1)–(2).)

The current global framework, Basel~III, requires banks to maintain minimum capital ratios and robust liquidity buffers that reflect their risk profiles. We benchmark stablecoin issuers against one key element of this framework: the simple leverage ratio, defined as equity divided by total assets.  This is the most basic Basel capital requirement and is the metric that U.S.\ regulators use as a trigger for ``prompt corrective action'' when banks become dangerously undercapitalized. Note that we are not arguing that it is appropriate for stablecoin issuers to meet commercial bank capital requirements; we find this exercise useful as a proxy for understanding a stablecoin issuer's capacity to withstand unexpected losses due to stress.

Using public filings, we construct a simplified leverage ratio for five prominent stablecoin issuers over three to ten recent quarters, depending on data availability. Data limitations prevent us from applying more complex risk-weighted capital ratios or the Basel liquidity standards, so our analysis focuses on this single, foundational measure. For banks, higher leverage ratios indicate greater loss-absorbing capacity and better protection for depositors. For stablecoin issuers, the leverage ratio captures the degree to which outstanding coins are overcollateralized. 

Let total assets equal the value of reserve assets and total liabilities equal the value of coins in circulation. Capital is then the difference between assets and coins, and the leverage ratio is calculated as follows:

  $$\text{Leverage ratio} \; = \; \frac{\text{assets}-\text{stablecoins}}{\text{assets}}$$

We apply this measure to five issuers that disclose sufficient information to approximate total assets and liabilities. The period we study is transitional: the reserve backing asset portfolios largely predate the GENIUS Act, and several issuers hold a broader range of assets than the Act would permit, including corporate bonds, secured loans, and volatile cryptoassets such as Bitcoin. Three issuers---Circle, PayPal, and Paxos---report reserve backing assets consisting primarily of cash or deposits and Treasury-related securities, and thus appear closer to the GENIUS standard, though we cannot verify full compliance. 

Under the Federal Deposit Insurance Corporation Improvement Act of 1991 (FDICIA), U.S.\ commercial banks must maintain leverage ratios of at least 4\% to be considered adequately capitalized and at least 5\% to be considered ``well capitalized.'' Banks with leverage ratios below 4\% are undercapitalized, below 3\% significantly undercapitalized, and below 2\% critically undercapitalized. Regulators are required to intervene and take ``prompt corrective action'' when banks fall below 4\% leverage ratios and, for critically undercapitalized banks, to move expeditiously toward receivership. See \citep{fdic2022enforcement} and \citep{gao2024bank}. Licensed stablecoin issuers currently operate without an analogous prompt-corrective-action regime.

Table~\ref{tab:leverage-ratio} reports the leverage ratios we compute.

\begin{table}[ht]
\centering
\caption{Leverage ratios (\%). Sources: \protect\citep{TetherISAE,CircleFinancials,PaxosPYUSD,PaxosTransparency,RippleTransparency}, also see Appendix~\ref{app:leverage-ratio-calculations} for methodology.}  
  \label{tab:leverage-ratio}
  \resizebox{\textwidth}{!}{%
  \begin{tabular}{lcccccccccc}
  \toprule
  \textbf{Stablecoin} & \multicolumn{4}{c}{\textbf{2023}} & \multicolumn{4}{c}{\textbf{2024}} & \multicolumn{2}{c}{\textbf{2025}} \\ \cmidrule(lr){2-11}
   & Q1 & Q2 & Q3 & Q4 & Q1 & Q2 & Q3 & Q4 & Q1 & Q2 \\ 
  \midrule
  Tether USDT & 2.99\% & 3.81\% & 3.71\% & 5.61\% & 5.68\% & 4.50\% & 4.86\% & 4.93\% & 3.75\% & 3.36\% \\
  Circle USDC & 0.16\% & 0.20\% & 0.21\% & 0.21\% & 0.16\% & 0.16\% & 0.18\% & 0.15\% & 0.04\% & 0.11\% \\
  PayPal PYUSD & -- & -- & 2.14\% & 2.29\% & 2.26\% & 2.16\% & 2.11\% & 2.16\% & 0.27\% & 0.42\% \\
  Paxos USDP & 0.00\% & -- & 0.00\% & -- & -- & 0.61\% & 0.01\% & 0.11\% & 0.13\% & 0.03\% \\
  Ripple USD & -- & -- & -- & -- & -- & -- & -- & 7.21\% & 4.43\% & 3.40\% \\
  \bottomrule
  \end{tabular}%
  }
  \end{table}

None of the three issuers whose reserves most closely resemble the GENIUS requirements---Circle, PayPal, and Paxos---would have been classified as adequately capitalized (under expectations for bank) over the period we study. Two would have been critically undercapitalized in every quarter for which data are available; the third moved from significantly to critically undercapitalized in the final two quarters.

In contrast, the issuers with riskier backing assets that are not GENIUS-aligned were better capitalized under the leverage ratio. Tether met the 4\% minimum leverage requirement in five quarters and exceeded the 5\% ``well capitalized'' threshold in two of those. However, these figures reflect reserve portfolios that included higher-risk assets such as corporate bonds and Bitcoin.\footnote{Please see Appendix~\ref{app:leverage-ratio-calculations} for a short discussion of the dependence of Tether's leverage ratio on Bitcoin. Tether has recently announced plans to issue a new U.S.\ dollar-based stablecoin, USAT, that it intends to be GENIUS-compliant. \citep{tether2025usat}} The remaining issuer, Ripple, met the minimum requirement in two quarters and was well capitalized in one of them.

Absent other mitigating factors, most stablecoin issuers in our sample would have been subject to prompt corrective action and required to raise additional capital or face increasingly stringent supervisory intervention had they been regulated as banks. The three issuers with reserve backing assets most closely aligned with the GENIUS standards might at times even have been candidates for receivership under FDICIA-style rules. In contrast, the remaining two with seemingly riskier backing assets generated higher leverage ratios that sometimes approached or met commercial bank requirements for this simplest measure of capital, suggesting that their balance sheets may be more resilient.   

Moreover, the dispersion of leverage ratios across all five issuers, ranging from near zero to just above 7\%, suggests that issuers either place little weight on capital as a risk mitigant or lack consensus on what constitutes adequate solvency. Some issuers seek indirect protection by investing heavily in overnight reverse repos, where borrowers post excess collateral under Fixed Income Clearing Corporation rules, but it is unclear whether such excess collateral would be sufficient or available to absorb large unexpected losses. In addition, retained earnings may grow over a longer time frame than what we studied and add to the capital buffer, but the growth of that capital element depends on an issuer's financial performance and decisions of its board and senior management.

Because the GENIUS Act appears to preclude applying commercial bank minimum leverage requirements directly to stablecoin issuers, policymakers may implicitly be relying on liquidity alone, rather than solvency, to ensure resilience. This makes robust liquidity-risk measurement and management crucial, which we address next.

\subsection{Comparing stablecoin issuers to MMFs}
\label{subsec:MMF's}

The liquidity structure of stablecoin reserve portfolios resembles that of money market mutual funds (MMFs). MMFs lack deposit insurance but are subject to regulations that limit their investments to mitigate run risk. MMFs invest in short-term, low-risk debt securities and seek to offer investors a cash-like instrument. Before 2014, MMFs maintained a fixed net asset value (NAV) of \$1 per share and were said to ``break the buck'' when their NAV fell below \$1. A 2014 reform required a switch to floating NAVs for prime funds, liquidity fees, and redemption gates to increase transparency and reduce run incentives \citep{ SEC_MMF2014}.

Anadu et al. \cite{anadu2024runs}  describe the close parallels between stablecoin issuers and MMFs: both raise funds by issuing liabilities payable on demand, and both aim to provide par-value redemption---MMFs in fixed dollars per share, stablecoin issuers in dollars or dollar-linked tokens. One key difference is that, since December 2015, MMFs have been allowed to lend to the Fed via its Overnight Reverse Repurchase Agreement (ON RRP) facility.\footnote{Prior to that date MMFs were eligible counterparties in limited amounts as part of a test program. See \citep{Hempel_ONRRP2023} for a discussion of the ON RRP.} This provides MMFs with a backstop that guarantees repayment at par with the dollar value of asset held. In terms of the hierarchy of money in Figure \ref{fig:Hierarchy_of_Money}, it effectively integrates government security MMFs into the same level as the banking system.\footnote{By lending directly to the Fed, MMFs avoid reliance on broker-dealer intermediation entirely for this portion of their liquidity management — the precise bottleneck that, as we show in Section 5.3, poses the greatest redemption risk for stablecoin issuers. Stablecoin issuers have no equivalent facility. Note, however, that the ON RRP facility does not shield MMF investors from a decline in the market price of securities held by the MMF. }

MMF regulation requires funds to hold a minimum volume of  short-duration assets. The Securities and Exchange Commission (SEC) limits eligible reserve assets to U.S. dollar-denominated securities and constrains portfolio maturity by requiring a dollar-weighted average maturity (WAM) of no more than 60 days and a weighted average life (WAL) of no more than 120 days. At least 25\% of assets must be held in instruments that mature within one business day (daily liquid assets) or five business days (weekly liquid assets). The referenced 2014 reform, which required prime institutional MMFs -- which invest in commercial paper -- to switch to a floating rate net asset value (``NAV''), triggered a massive exit of 70\% from institutional prime funds into Government MMF's (approx. \$825bn) when it came into effect in 2016, the majority of which shifted into fixed NAV government securities funds \citep{ALLEN-MMFS}\footnote{\citet{Holmstrom2015} describes a MMF with a fixed NAV as ``informationally insensitive''. The 2016 shift of prime funds to floating NAV made them less money-like. This has been proposed as an explanation for the magnitude of withdrawals \citep{ALLEN-MMFS}}. See Section \ref{subsec:information} for a more detailed discussion of informationally insensitive securites applied to the Treasury markets.

We adapt the Daily Liquid Assets (DLA) and Weekly Liquid Assets (WLA) concepts to stablecoin issuers. Before the GENIUS Act, issuers varied widely in the detail and quality of their disclosures. Circle and Paxos provide granular audits and attestation reports, including CUSIP-level detail, which allow reasonably precise estimates of DLA and WLA shares. Tether, by contrast, discloses the composition of its USDT reserves but not maturities, making strict application of the framework impossible. For the issuers where we can approximate DLA and WLA, a clear pattern emerges: reserves have migrated toward highly liquid securities---U.S.\ Treasuries and overnight repos---leaving DLA and WLA well above the MMF thresholds. At least some issuers thus appear to be converging, intentionally or not, toward MMF-style reserve practices.

Not all issuers have followed this path. Some continue to hold assets with potentially higher price volatility, including corporate bonds, precious metals, and cryptocurrencies such as Bitcoin. Stablecoins backed by such portfolios may face a greater risk of ``breaking the buck'' if many \stc owners seek to redeem simultaneously.

The 2014 reforms required MMFs to impose liquidity fees and redemption gates during periods of market stress. These mechanisms were designed to slow withdrawals and reduce investors' incentives to run preemptively. However, subsequent analysis by the Government Accountability Office (GAO) finds that these reforms did not prevent large-scale redemptions from MMFs during the COVID-19 pandemic. Indeed, investors may have accelerated withdrawals to avoid the possibility of future gates or fees \citep{gao2023moneymkt}. In reponse to this concern, in July 2023 the SEC abolished discretionary redemption gates for institutional prime and tax-exempt MMFs entirely, replacing them with a mandatory liquidity fee triggered when net redemptions exceed 5 percent of net assets in a single day \citep{SEC2023MMF}. 

\subsection{Summary and implications}

\textbf{Comparative resilience of banks and MMFs}. Taken together, the five balance sheet characteristics—the share of liabilities redeemable on demand, the liquidity of assets, control over balance sheet growth, access to public support, and capitalization—suggest that commercial banks enjoy a generally more resilient structure that can absorb unexpected losses and thus maintain par-value claims over the long run than do current stablecoin issuers. Banks limit the fraction of immediately redeemable liabilities, exercise greater control over balance sheet evolution, hold larger capital buffers, and benefit from deposit insurance and central bank backstops. Nonetheless, some bank assets, like loans, are generally less liquid than stablecoin assets such as Treasury-related securities and cash held in commercial bank accounts, suggesting that stablecoin issuers may be better able to service unexpected redemption requests in the short run.

With regard to liquidity, MMFs have been specifically designed with guardrails to mitigate run-risk and over time some stablecoin issuers have developed reserves that tend to mimic MMF asset composition.  Stablecoin issuers that mirror MMF asset portfolios likely have  similar liquidity to MMFs, from purely an asset perspective; however, the GENIUS Act states that redemptions for stablecoin issuers must occur "without delay." Although the effectiveness of redemption gates and liquidity fees have been mixed, stablecoin issuers lack access to similar tools to potentially slow down redemptions during runs.


\textbf{Risks for stablecoin issuers.} Stablecoin issuers typically invest in high-quality, liquid assets such as Treasury-related securities, but most of the issuers in our sample appear to have thin capital buffers by bank standards. The most GENIUS-aligned issuers had the weakest capitalization when measured using the leverage ratio, suggesting that their balance sheets are the most fragile. Our findings indicate that, in the face of large interest-rate, operational, or market shocks, many issuers could become insolvent, jeopardizing their ability to honor par redemptions. Moreover, if many issuers were forced to sell Treasury-related securities simultaneously, the resulting market impact could itself impair liquidity and valuations, a possibility we explore in Section~\ref{sec:redemption_risk}. 

\textbf{Regulatory implications.} Because the GENIUS Act does not appear to authorize the straightforward application of Basel-style capital requirements to preserve solvency, regulators will need to rely on other tools to strengthen the resilience of stablecoin balance sheets. One natural candidate is a framework of explicit liquidity requirements, adapted to the structure of stablecoin issuers and informed by experience with MMFs.


\section{Redemption risks due to Treasury market fragility}
\label{sec:redemption_risk}

GENIUS limits \si asset allocation to bank deposits and Treasury securities. The latter requires a \si to liquidate Treasury securities to meet its redemption payment obligations. Though these are large, liquid markets, they have suffered disruption in the past. 
%
In this section we review two recent episodes of disruption of the Treasury markets that were associated with relatively small shocks to the demand for Treasuries. We explain how this fragility is amplified by the interaction of accounting rules and bank capital regulations, which limit the volume of repo trades a bank affiliate broker-dealer can intermediate, and a mal-distribution of central bank reserves that impedes the channeling of reserves to broker-dealers to meet an uptick in selling pressure. These factors create a two-way channel between Treasury market stress and redemption runs. In one direction a run to redeem that increases selling pressure -- even by a modest amount -- can disrupt the market. In the other direction, a modest disruption in pricing or processing of sales can trigger a run to redeem. We then argue that the Fed's standing repo facility does not achieve the intended goal of alleviating the capacity constraint in the Treasury market. 
We discuss the effects of granting \si{s} access to the Fed's balance-sheet and point out a crucial policy dilemma: On the one hand, redemption risk can be mitigated by allowing \si{s} to hold central bank reserves. On the other hand, granting \si{s} access to the Fed's balance sheet could alter the transmission of monetary policy and disintermediate banks.

\subsection{Reliance on the Treasury markets}
\label{subsec:fragility}

The largest markets for Treasury securities are (i) the secondary market for sale and purchase of Treasuries and (ii) the repo market.\footnote{Original issue auctions of Treasuries is a third market that can only be accessed by Primary Dealers and individual investors.} The former has daily turnover of approximately \$900 billion \citep{Liang2025}. The latter has average daily turnover of approximately \$12 trillion \citep{HempelKahnShephard2025}. These are the two largest volume financial markets in the world. By conventional metrics used in finance  they are the most liquid markets and by historical price variability they are the most stable. Currently, \si backing assets are concentrated in short term Treasury securities, which provide a cushion against capital loss (Section \ref{sec:insolvency}). Nevertheless, there is a fault line running through the Treasury market that makes it fragile in the face of a spike in selling pressure, such as can occur during a run to redeem a \stc.

A redemption that is met  by liquidating treasuries triggers a two-way interaction with the market that can amplify stress. In one direction, an uptick in redemption demand can push broker-dealers to the limit of their capacity to purchase Treasuries. In the opposite direction, when broker-dealers have reached that limit---possibly for reasons unrelated to redemptions---the disruption to processing redemptions can cause a run to redeem. We examine those interactions and show that the underlying source of redemption risk arises from the fact that Treasury markets are intermediated by broker-dealers with limited capacity to process trades, and in fact a reduced incentive to process trades, in times of market stress. 

\subsection{Recent disruptions in Treasury markets}
\label{subsec:recent_disruptions}

Table~\ref{tab:yield-volume-events} summarizes two recent instances of disruption in the Treasury markets; one in the repo market and one in the secondary market.

\begin{table}[H]
\centering
\small
\begin{adjustbox}{max width=\textwidth}
\begin{tabularx}{\textwidth}{@{}p{3.1cm} p{2.5cm} X X@{}}
\toprule
\textbf{Event} & \textbf{Date / window} & \textbf{Yield dislocation (Treasuries / repo)} & \textbf{Trading volume effect (Treasuries / repo)} \\
\midrule

September 2019 repo spike &
16--18 Sept.\ 2019 &
SOFR jumped from about 2.43\% on Sept.\ 16 to 5.25\% on Sept.\ 17, with some intraday transactions reportedly near 10\%; the spread of SOFR over the top of the federal funds target range exceeded 300 basis points \citep{anbil2020money,ofr2023repo}. &
SOFR transaction volume on Sept.\ 17 reached about \$1.18 trillion, around \$20 billion higher than the previous day (a low single-digit percent increase), indicating that the episode was driven more by an extreme funding-rate dislocation than by an unusually large surge in underlying transaction volume \citep{anbil2020money}. \\[0.5ex]

March 2020 ``dash for cash'' in Treasuries &
Late Feb.--late Mar.\ 2020 (esp.\ week ending 4 Mar.) &
The 10-year Treasury yield fell to an intraday low near 0.5\% and then backed up by roughly 60+ basis points over a few days; off-the-run issues underperformed on-the-run benchmarks and bid--ask spreads widened by an order of magnitude, reflecting severe dysfunction in Treasury market liquidity \citep{fleming2020treasury,haddad2021when}. The dislocation was maturity-segmented: investors shed longer-duration notes and bonds while simultaneously seeking safety in short-term Treasury bills, driving bill yields sharply lower even as longer yields rose \citep{VISSINGJORGENSEN202119,HeNagelSong}.
&
Overall Treasury market trading volume reached record highs: in the week ending March 4, average daily volume exceeded \$1 trillion---roughly twice its post-crisis average---implying on the order of a 100\% increase in cash-market turnover relative to normal conditions; primary-dealer volumes in mid-March were similarly elevated \citep{fleming2020treasury,nbtf2021ust}. \\

\bottomrule
\end{tabularx}
\end{adjustbox}
\caption{Yield dislocations and associated trading volume increases in U.S.\ Treasury and Treasury repo markets during two recent stress events: the September 2019 repo spike and the March 2020 ``dash for cash''.}
\label{tab:yield-volume-events}
\end{table}

\begin{figure}[H]
\begin{center}
\includegraphics[page=1,width=0.6\textwidth,height = 0.12\textheight]{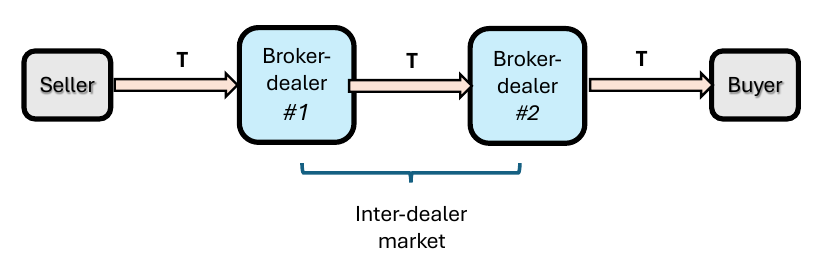}
\end{center}
\caption{Treasury market transaction chain}
\label{fig:treasuries_chain}
\end{figure}

The yield spike in September 2019 was not associated with any appreciable change in transaction volume. March 2020 is more relevant for our analysis, since it was characterized by a run to exchange Treasuries for cash. In that event trading volume increased by an estimated \$500 billion.\footnote{Secondary Treasuries sales increased from a weekly average of \$2.5-\$3 trillion for four weeks prior to the event, to \$4.9 trillion the week of the event. Source: TRACE data \citep{NBTF_2021_US_Treasury_Market_COVID} and \citep{IAWG_2021_Staff_Progress_Report}} Note, however, that this represents  \textit{gross} volume. An important fact is that the secondary Treasury market is intermediated by broker-dealers, who stand between the ultimate seller and buyer of Treasuries.\footnote{``Traditionally, dealers have been the predominant market makers, buying and selling securities from customers to meet customer trading needs'' \citep{Fleming_Treasdealers}.} Figure \ref{fig:treasuries_chain} displays a chain of trades in the Treasury market with ultimate seller (or repo borrower) and buyer (or repo lender) at each end and two broker-dealers intermediating in the middle, which represents the average length of trading chain in the two Treasury markets.\footnote{A chain with two broker-dealers is the most frequently occurring chain type in the US Treasury repo market \citep{AronoffOFR}.} The liquidation starts with the sale of Treasuries to broker-dealer \#1. The Treasuries are either absorbed by a broker-dealer, or are passed on to an ultimate buyer on the right. 

We estimate the volume of liquidated Treasuries in the March 2020 event by sperading the transactions across the inter-dealer market and adjusting for retention by broker-dealers. This amount was \$70-\$75 billion \citep{FlemingEtAl2022}.\footnote{Fleming et al. Chart 13 is a visual representation of confidential FR 2004A Weekly Reporting of Dealer Positions. The value range is based on visual inspection.} Filling in with median reported numbers\footnote{We checked the accuracy of our estimates as follows. The TRACE data show an approximate aggregate daily increase in Treasuries transactions of \$500bn during the week of the ``dash for cash'', March 16-20, 2020, relative the average volume during the preceding four weeks. The TRACE data counts every transaction. We divide by 3 to obtain an estimate of net Seller sales. The resulting amount is consistent with the TRACE volume for customer (i.e., non broker-dealer) sales reported in Chart 11 of~\citet{FlemingEtAl2022}. This coincidence shows that the chain length in Figure \ref{fig:treasuries_chain} is consistent with the data.}  yields:

\[ \text{Total volume} = \underset{\textbf{{\color{blue}seller volume}}}{\underbrace{{\color{blue}\$216bn}}} + \underset{\text{broker-dealer retention}}{\underbrace{\$72.5bn}} + \underset{\text{inter-dealer volume}}{\underbrace{\$141bn}} + \underset{\text{buyer volume}}{\underbrace{\$141bn}} =  \$500bn\]

This indicates that the Treasury market meltdown of March 2020, which was the most severe disruption in decades, involved a selloff of just over \$200 billion. This represents less than one percent of Treasuries held by the public. It underscores the extreme fragility of the Treasury market and explains the subsequent focus by academics, industry participants and regulators to develop proposals to improve Treasury market resiliency.  It is approximately 2/3 of the current market value of \stc{s}.\footnote{A redemption run will cause an increase in notional sales for Treasuries across the maturity spectrum. This results from a liquidation of a repo position where the \si elects not to "roll over" into a new repo after selling the Treasury at the second-leg. When that occurs the counterparty must find another way to fund its Treasuries---either a repo with a new counterparty (if one can be found) or a sale. The sale will involve Treasuries across the maturity spectrum, which is the composition of repo collateral.}

The projected rapid growth in \stc{s}, if it materializes, will generate heightened risk of Treasury market disruption in the future. Optimistic forecasts of \stc adoption predict a market value of \$2-\$4 trillion by 2030 \citep{Citi_2025_Stablecoins2030}, which is a nontrivial portion of the \$26-\$27 trillion of the publicly held US government debt currently forecasted for 2030 \citep{CBO_stcproj}. On the evidence of recent disruptions, even a modest uptick in net redemption in the future could dislocate the Treasury market and compromise the ability of \si{s} to maintain \pv.\footnote{A recent study estimates a 2.5-5 basis point decline in T-bill yields following a \$3.5 billion inflow into stablecoins, which suggests that Treasuries are highly sensitive to \stc redemptions \citep{ahmed2025stablecoins}.}

\subsubsection{Interaction of duration and stablecoin issuer backing assets}

A feature of the March 2020 episode with direct bearing on stablecoin redemption risk is that selling pressure was concentrated in longer-duration, off-the-run securities, while demand for short-term instruments remained comparatively firm as investors sought safety within the Treasury market itself \citep{VISSINGJORGENSEN202119}. In the counterfactual where \si{s} were operating during the episode,  maturity segmentation has asymmetric implications for stablecoin issuers depending on the composition of their backing assets.

To the extent a stablecoin issuer meets redemption demand by declining to roll over maturing repo positions, the burden of liquidation falls on its repo counterparties, who must either secure replacement financing in an already stressed market or sell their Treasury collateral outright. That collateral is predominantly longer-duration, off-the-run notes and bonds -- in the tri-party repo market, coupon-bearing securities routinely account for roughly 70–80 percent of posted collateral \citep{Copeland2012,Paddrik2021TripartyRepo}. A non-rollover by stablecoin issuers during a redemption run could therefore amplify the off-the-run selloff that characterized March 2020, exacerbating the very dislocation that threatens their own ability to maintain par-value.

By contrast, direct sales of short-duration Treasuries — which under GENIUS are limited to maturities of 93 days or less -- would likely occur at elevated prices given strong flight-to-safety demand for short-dated instruments, providing some insulation for the stablecoin issuer's own solvency. The systemic risk, however, runs through the counterparty channel described above, by which a stablecoin redemption run can propagate stress into the longer-duration segment already under pressure \citep{HeNagelSong}.\footnote{The GENIUS Act's 93-day maturity limit governs the stablecoin issuer's direct Treasury holdings, not the maturity of collateral posted by repo counterparties. In a reverse repo, the stablecoin issuer is the cash lender; the borrower determines the collateral composition. As documented in \cite{Copeland2012} and \citep{Paddrik2021TripartyRepo}, tri-party repo collateral is predominantly longer-duration, off-the-run coupon securities. A stablecoin issuer's short-duration direct portfolio therefore does not shield the broader market from the longer-duration collateral liquidation that a non-rollover forces upon its counterparties.}

\subsection{Bottlenecks in the Treasury markets}
\label{subsec:bottlenecks}

Bottlenecks arise from the intermediated structure of the two Treasury markets. The source of bottleneck in each market is caused by a balance-sheet constraint and a limitation on the volume of central bank reserves available to broker-dealer affiliates of banks.

\textbf{Secondary Treasury market: limitations on central bank reserves.} There are two ways a bank affiliate of a broker-dealer can facilitate the purchase of a Treasury. If the payment is made to a seller account at the subject bank, it increases the seller's deposit account and increases its assets by the value of the acquired Treasury. This pushes its leverage ratio closer to the regulatory limit (see below). If the seller holds a deposit account at another bank, the subject bank is required to send central bank reserves of equal value to the seller's bank. This requires the bank affiliate of the broker-dealer to have adequate reserves on hand. If it runs short of reserves it cannot acquire the Treasury.

\textbf{Treasuries repo market: limitations on balance-sheet capacity.}  A broker-dealer acting as repo lender must transfer a \bd at the first-leg. This can be accomplished in either of the two ways discussed above, with the same consequences. An additional limitation on intermediation capacity in the repo market arises from the interaction of accounting rules and bank capital requirements. Accounting rules require a broker-dealer to increase recorded assets by approximately the value of the intermediated repo trade, while bank capital regulations limit the permissible increase in assets. An increase in the volume of repo intermediation will move broker-dealer assets closer to the regulatory limit (see below).\footnote{Repo trades are treated as secured financing under GAAP rules (See FASB ASC 860-30-25-2 and ASC 860-10-40-5). The first-leg sale of Treasuries causes recorded assets to increase by the sale price. Intermediation involves the first-leg purchase and sale of a Treasury. Therefore, intermediation causes recorded assets to increase. For an example application of the repo accounting rule see \citet{Salerno-Repo-Accounting}. For a detailed explanation of repo accounting rules see \citet{AronoffrepoBS2025}.}

\subsubsection{The supplementary leverage ratio constrains repo intermediation}

The supplementary leverage ratio (``SLR'') is the bank capital regulation that has imposed the tightest constraint on the ability of banks to allow their broker-dealer affiliates to increase repo intermediation volume during stress events. The SLR places a lower bound, denoted by $\underline{L}$, on the ratio of bank capital to unweighted assets plus off balance-sheet exposures, of 3\% with an additional 2\% for large globally systemically important banks (GSIBs)~\citep{BIS2010}.\footnote{The exposures are mostly related to derivatives positions.} 

\begin{center}
SLR: capital/(assets + exposures) $\geq$ 3\% + 2\% for GSIBs = $\underline{L}$
\end{center}

In recent years the largest dealer banks have been operating near their SLR lower bound, which limits their capacity to absorb an increase in repo volume (Table \ref{tab:slr-gsibs-q3-2025}).

\begin{table}[H]
\centering
\begin{tabular}{@{}l c@{}}
\toprule
\textbf{Bank} & \textbf{SLR, Q3 2025} \\ 
\midrule
JPMorgan Chase    & 5.8\% \\
Bank of America   & 5.8\% \\
Citigroup         & 5.5\% \\
Goldman Sachs     & 5.2\% \\
Morgan Stanley    & 5.5\% \\
\bottomrule
\end{tabular}
\caption{Supplementary Leverage Ratio (SLR) for Major U.S. GSIB Dealer Banks, Q3 2025\\ Source: Corporate 10-Q filings.}
\label{tab:slr-gsibs-q3-2025}
\end{table}

The combination of the extremely large volume of recorded assets required to intermediate the \$12 trillion repo market, the low profit margin earned from repo intermediation and the proximity to the SLR lower bound of the major repo intermediaries has prompted concern that the SLR regulation has placed a binding constraint which is limiting the capacity of their broker-dealer affiliates to intermediate the US Treasuries cash and repo markets. In his 2017 Baffi Lecture economist Darrell Duffie expressed concern over the restrictive impact of the SLR on intermediation:
\begin{quote}
The concern is instead that the amount of intermediation provided by banks to low-risk asset markets has become inefficiently low... the largest U.S. dealer banks must carefully consider the impact of the leverage ratio rule (SLR) on their minimum capital levels when deciding how much of their balance-sheet to allocate to safe asset intermediation \citep{Duffie2017}, Chapter 2).
\end{quote}

\subsubsection{The standing repo facility does not relieve balance-sheet constraints to Treasuries intermediation}
\label{subsubsec:srf}

A feature of the September 2019 repo disruption and the March 2020 secondary market disruptions was the inability of the bank affiliates of broker-dealers to acquire central bank reserves to meet demand. This is superficially striking because of the abundant central bank reserves in the banking system at the time. The problem has been identified as a mal-distribution of reserves, whereby banks with excess reserves could not send those reserves to broker-dealers because there was no active inter-bank lending market and the banks holding excess reserves did not intermediate in the Treasury markets \citep{copeland2021reserves, afonso2021market}. The SRF is intended to remedy this by enabling broker-dealers to access reserves directly from the Fed.  The SRF enables a bank to sell Treasuries to the Fed at the first-leg in exchange  for the reserves it requires to fund the purchase of Treasuries from -- in this example -- the \si. The bank can thereafter roll over the repo at the second-leg until it finds a buyer for the Treasuries it purchased from the \si. The objective of the SRF, as stated by the Fed, is to avert a meltdown by providing liquidity to broker-dealers to enable them to absorb a surge in selling by market participants \citep{FedPowell,FedSRF}.

\begin{figure}[H]
\begin{center}
\includegraphics[page=1,width=0.7\textwidth,height = 0.2\textheight]{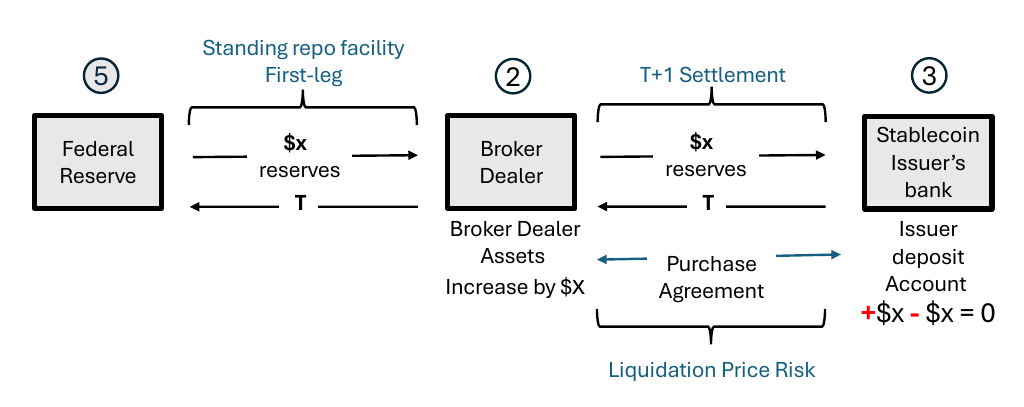}
\end{center}
\caption{Stablecoin redemption process}
\label{fig:redemption_process}
\end{figure}

Figure \ref{fig:redemption_process} modifies the redemption process depicted in Figure \ref{fig:redemptionstc} by adding a Step 5 where the Treasury buyer -- shown as a broker-dealer -- obtains reserves from the SRF, which it uses to complete its purchase of the security from the \si. 

When the \si employs a redemption intermediary, such as Circle uses Coinbase and several other firms, these market-makers add another layer to the redemption process. For example Circle would need to liquidate securities -- which may not be Treasuries -- to acquire the \stc from the redeemer. After that, the process is repeated when Coinbase redeems with Circle. It is unclear how the redemption intermediary may affect the demand for central bank reserves. It has a neutral effect when it uses \bd{s} to acquire \stc{s} and then retains the \stc during the stress period. It magnifies demand when it liquidates securities and then immediately redeems with the \si.

Contrary to the stated intent of the SRF, we believe its significance in alleviating market stress is limited. The key thing to note is that in Step 2 the broker-dealer enters into a reverse-repo, which requires it to increase its assets by the amount of reserves it borrows from the Fed. A prohibition on centrally clearing repo trades with the Fed prevents the broker-dealer from netting the reverse-repo with an un-netted repo, to the extent it has un-netted repo volume.\footnote{The Fed is excluded from the central clearing mandate \citep{SEC-UST-CCP-Rule-2023}. Central clearing novates initial trades contracts and replaces them with a central clearing counterparty (``CCP'') as the counterparty for each trade. This ensures that trades are multilaterally netted.} An analysis by \citet{BowmanFedNotes} concludes that the central clearing mandate, which requires broker-dealers to centrally clear all Treasuries repo and secondary trades (excluding trades with the Fed), will not significantly reduce broker-dealer recorded assets and therefore will not significantly affect their SLR.\footnote{``Combining a detailed analysis of the rules involved in calculating the SLR with a unique set of regulatory data, we conclude that expanding central clearing would have relatively limited effects on the level of SLRs.''} 
This limits the volume of reserves that broker-dealers can obtain from the SRF, even after the central clearing mandate is implemented.\footnote{Recall that repo intermediation increases recorded assets, which pushes the SLR closer to its lower bound.}  Finally, a November 2025 reform to the SLR effectively reduces the lower bound \citep{eslr_rule_2025}, however it is likely to only marginally increase broker-dealer intermediation capacity.\footnote{A 'back-of-envelope' calculation predicts an approximate \$500 billion increase in repo intermediation capacity. The figure is arrived at by estimating an approximate increase of \$1 trillion in balance-sheet capacity at each of the five largest broker-dealer bank affiliates, multiplied by 10\%, which is an upper bound to the Q4 2025 allocation of assets to repo. This amounts to less than a 5\% increase in the \$12 trillion Treasury repo market. Notably, the intermediated structure of the repo market implies that the attainable increase in volume of net repos is a fraction of \$500 billion total increase. }  

Our main conclusion is that the interaction between accounting rules and bank capital regulations, at a time when broker-dealer affiliate banks are operating near their SLR lower bounds, limits the volume of reserves that broker-dealers can obtain from the SRF during a period of market stress, such as when \si{s} are required to liquidate a large volume of Treasuries to maintain \pv. This presents a tangible risk that \si{s} will be unable to meet the \pv requirement at all times.

\subsection{Informationally insensitive securities and the risk of stablecoin runs}
\label{subsec:information}

Here we discuss the other channel of causation; where a modest disruption in the Treasury markets can induce a run to redeem. \citet{Holmstrom2015} and \citet{Dang_banks} posit two interdependent conditions that a security must fulfill to be accepted as a safe asset: (1) that it credibly maintains par-value with central bank money or some other reference asset and (2) that agents do not---and cannot without incurring cost---verify the underlying condition of the assets backing the security. (1) is self-evident. (2) is subtle. Bank deposits satisfy (2) because depositors do not observe the granular composition of loans that comprise the majority of bank assets, or its off balance-sheet derivatives and interest rate exposures. It means that depositors do not know the underlying condition of each loan and exposure. When confidence in the bank’s solvency is high, no depositor has an incentive to try to gain an advantage by spending to become better informed about the bank’s asset portfolio. But when solvency is questioned, confidence quickly evaporates, as occurred in  2023 when SVB’s losses on its Treasuries portfolio was revealed and rendered it insolvent.  

\begin{quote}

Panics happen when information-insensitive debt turns into information-sensitive debt… A regime shift occurs from a state where no one feels the need to ask detailed questions … to a state where there is enough uncertainty that some investors begin to ask questions about the underlying collateral … This can lead to reduced liquidity and rapid drops in prices. \citep{Holmstrom2015}
\end{quote}

Another  example of this phenomenon occurred in 2008 when the Reserve Primary Fund, at the time the largest MMF, “broke the buck” by falling below par (with the dollar) by a mere 3\%, which led to mass redemptions and a collapse of the fund. \citet{anadu2024runs} documented tx`hat both MMFs and \stc{s} have experienced runs in response to small deviations below par.\footnote{Par in the context of a MMF refers to  fixed \$1.00 per share price.} Perhaps the most striking example in recent years was the shift out of prime MMF's in 2016 as discussed in Section \ref{subsec:MMF's} when prime funds were required to adopt floating NAV's. The rule change transformed prime MMF's from an informationally insensitive security to information sensitive debt. This historical causal pattern can be understood by considering the role of broker-dealers in intermediating the Treasury markets (as discussed above). We have seen that a redemption involves a sale of Treasuries to a broker-dealer (to the extent the \si does not have \bd{s} on-hand to cover the redemption), and that broker-dealer capacity to purchase Treasuries is constrained by the limitation on the ability of bank affiliates of broker-dealers to obtain supplemental reserves from other banks or from the SRF. An additional factor is that the remaining capacity of a broker-dealer to purchase Treasuries involves elements that are not visible to outsiders, such as the ability of the affiliate bank to re-allocate reserves to the broker-dealer by making adjustments in other parts of its balance-sheet (``Banks are opaque by design'' (Dang et al. \cite{Dang_banks})). This opaqueness makes  the capacity of broker-dealer intermediaries in the Treasury markets to process transactions informationally insensitive. At most times, \stc holders reasonably assume that redemptions can be processed. However, a small disruption in \pv of the \stc can signal that broker-dealers have reached the limit of their capacity to facilitate Treasuries purchases, and cause a regime shift where parties lose confidence in the ongoing ability of the \si to effectuate redemptions at par. This, in turn, could trigger a  redemption run. Aldasoro et.al. \cite{aldasoro2023onpar} point out that markets where \stc{s} are traded do not have frictions that would relieve pressure to redeem, which make \stc{s} a highly runnable asset.\footnote{Aldasoro et.al. \cite{aldasoro2023onpar} compare \stc{s} to Eurodollars. A comparison to MMF's would show \stc{s} to have a higher run-risk because a MMF can ensure its ability to meet redemptions by holding reverse repos with the Fed.} 

\subsection{Stablecoin issuer access to Fed reserves: a solution with complications}
\label{subsec:dilemmas}

\textbf{A solution:} Granting \si{s} access to the Fed discount window or the SRF would improve the ability of a \si to meet redemption requests during a run.\footnote{Moreover, the existence of such access will reduce the incentive of agents to participate in a run by making them less fearful that a run will threaten the stability of \pv.} It would do so by enabling the \si to borrow against Treasuries directly, thereby avoiding a market bottleneck. Direct access is required because, as we explained in Section \ref{subsec:bottlenecks}, 
 the liquidity provided by the standing repo facility to Treasury market intermediaries does not reliably flow through to provide liquidity to \si{s}. \footnote{A recently proposed additional source of liquidity that could serve as a backing asset, called PORTS, involves issuance of overnight Treasuries that settle on-chain  \citep{Duffie-PORTS}. PORTS would reduce the interest rate risk of holding Treasuries. However, the required daily auction of Treasuries would be subject to the bottlenecks discussed above, loosened to the extent that PORTS enables broker-dealers to make more efficient use of their Treasury collateral. Another implication is that PORTS may reduce demand for \stc{s}, since it would be an alternative on-chain safe asset.}



\textbf{Complications:} There are two cases to consider.

\begin{itemize}
\item  (Discount window/ SRF access.) There is no doubt that every firm and household would individually benefit from access to central bank liquidity. However, universal access would  create soft budget constraints throughout the economy, which would undermine efficiency \citep{kornai2003understanding}. The question is whether the Fed should view \stc{s} as a form of money for which it has responsibility to provide support. Notably, a condition for direct access would require that \si{s}  become subject to bank-type regulations. This could substantively increase operating cost and place into question the solvency of \si{s} in a low interest rate environment.

\item (No direct access.) An alternative is to allow \si{s} to have ``Fed skinny accounts'', as recently described by Fed Governor Waller \citep{Waller20251021}. While the concept is still speculative, the essential idea is to allow \si{s} to hold non-interest paying central bank reserves without directly transacting with the Fed. On one hand, this would reduce cost and complications involved in redemption and minting transactions. The \si would not be reliant on a third-party bank to intermediate, replacing the \si{'s} bank with the \si in Figures \ref{fig:redemptionstc} and \ref{fig:minting_process}.  On the other hand, this would not solve the bottleneck discussed in Section \ref{subsec:bottlenecks}, except to the extent that the \si could hold reserves as a backing asset. It would still have to liquidate Treasuries through market intermediaries.\footnote{
One approach to enable nonbanks to use central bank reserves to support payments is Fnality. Under this proposal, a firm would deposit a security with the central bank in exchange for use of reserves as collateral for on-chain payments. The Bank of England has designated its Sterling Fnality Payment System (£FnPS) for Settlement Finality protections \citep{Fnality_BoE}.}

Granting nonbank institutions access to reserves adds a new channel of  monetary transmission, which increases the complexity of conducting monetary policy. The marginal effect of diverting reserves into \si portfolios contracts reserves and deposits in the banking system, because the \si acquires reserves by sending \bd{s} to the Fed, which is accompanied by receipt of an equal value of reserves sent from the deposit issuing bank. The Fed can offset the loss of reserves held by banks through open market purchases of securities. This may serve the objective of neutralizing the effects on bank balance-sheets. However, it may conflict with other policy objectives. For example, the open market operation may push short term interest rates below the Fed funds target range; or the expansion of reserves may create inflationary pressure or conflict with a Fed goal of reducing the size of its balance-sheet.  An exploration of these matters lies beyond the scope of this paper. Our point is to highlight that granting \si{s} access to central bank reserves raises issues that merit attention. 



\end{itemize}

\subsection{Summary}

Stablecoin issuers are minimizing risk of capital loss and maximizing flexibility to meet redemptions by investing their backing assets in short duration securities (Section \ref{sec:insolvency}). Their asset portfolios are more conservative that commercial banks by a long-shot. And yet, they are at much greater risk of defaulting on their monetary liabilities compared to banks. The difference is that banks are integrated into the Fed’s monetary system and have direct access to the Fed’s balance-sheet in a stress scenario. Stablecoin issuers, by contrast, are exposed to the risk of market bottlenecks which are likely to arise in stressed situations. One solution to  the liquidity risk faced by \si{s} is to enable them to borrow reserves from the Fed. However, allowing \si{s} to hold central bank reserves could, depending on the rules, increase costs due to regulatory oversight, disintermediate banks, and reduce lending in the economy.


\section{Technical and operational risks}
\label{sec:tech-and-op-risks}


In this section, we explore risks associated with the operation of a \stc payment system. The use of a blockchain platform introduces risks that are not present in the incumbent financial system, which is operated on platforms controlled by (central and commercial) banks. We first compare features of a decentralized blockchain with a centralized network of intra and inter-bank payments. We then enumerate technical and operational risks associated with the blockchain itself and its interaction with the smart-contract that holds the \stc accounts and executes transfers. After that, we discuss measures that can be taken to mitigate these risks.

\subsection{Centralized versus decentralized payment platforms}

\textbf{Consensus on transaction finality.} A core difference between decentralized blockchains used for stablecoin payments and incumbent bank-operated payment rails lies in the type of ``consensus'' required to finalize transactions. On a public blockchain, transaction finality is an outcome of a protocol-level consensus mechanism (e.g., proof-of-stake or proof-of-work), in which a distributed set of validators orders transactions into blocks and the network converges on a canonical ledger. Finality may be probabilistic (strengthening with additional confirmations) or economically deterministic once a finality rule is met, but it is ultimately anchored in software rules and incentive compatibility among independent participants. By contrast, bank payment systems rely on institutional trust and governance rather than computational consensus: transactions are finalized when they are accepted under the operating rules of the relevant payment system and settled across bank accounts, often in central bank money. In this setting, settlement finality is primarily a legal and operational concept---backed by supervisory oversight, access controls, and well-defined procedures for exception handling---rather than an emergent property of open-network coordination.

\textbf{Recourse.} These different finality models map directly into differences in responsibility for system operation and the availability of legal recourse when payments are mishandled. In centralized bank platforms, responsibility is clearly allocated: regulated banks, payment system operators, and (for settlement in reserves) the central bank each have defined operational obligations, supervisory expectations, and liability regimes. Errors can often be investigated, reversed, or compensated through established dispute-resolution channels, and regulators can demand remediation through examinations and enforcement actions. In decentralized \stc payments, operational responsibility is more diffuse. Validators provide the ordering service but typically do not have a customer relationship with end users; smart contracts may execute transfers automatically with limited scope for ex-post intervention; and failures can arise from software vulnerabilities, governance attacks, key compromises, or outages at intermediaries such as wallets and exchanges.

\textbf{Record-keeping.} These differ in ways that matter for transparency, auditability, privacy, and supervisory access. In a stablecoin system implemented on a public blockchain, the ledger of token ownership and transfers is typically maintained as a shared, append-only record, and the issuer's liability to token holders is represented by balances in a smart contract. This can make transactional histories widely observable, machine-verifiable, and reconciled by design. In incumbent bank payments, by contrast, customer deposits and intrabank transfers are recorded on bespoke internal ledgers maintained by each institution, while interbank payments are reconciled through centralized messaging and settlement arrangements; the central bank and payment system operator maintain authoritative records of reserve balances and the settlement positions of participating banks. The result is a layered accounting architecture that is less publicly transparent than a blockchain ledger and requires costly reconciliation, but embeds clear permissioning, confidentiality, and supervisory controls.

\subsection{Stablecoin infrastructure risks}

\textbf{Incentive risks.} Many \stc{s} rely on public proof-of-stake (PoS) blockchains where validators are rewarded (fees and token issuance) for producing blocks and can be penalized (``slashed'') for violating protocol rules. For regulators, the key point is that transaction finality and throughput are delivered by an incentive system rather than by a supervised operator with an explicit contractual responsibility to provide neutral, continuous service. Validators typically have discretion over transaction ordering and can prioritize transactions that maximize revenue, including through maximal extractable value (MEV). In normal conditions, this can affect fairness and predictability for payment users; in stress, coordinated delay or censorship can occur and produce ``liveness'' failures in which the network remains online but does not reliably process transactions. These dynamics can impair payments without a clear liability framework or guaranteed remediation pathway comparable to those in regulated payment systems.

\textbf{Transaction processing risks.} A \stc might not be able to maintain \pv if users cannot reliably transfer tokens or if the \si cannot reliably mint and redeem. On-chain, these functions are simply transactions; if the underlying network is congested, attacked, or otherwise unable to process transactions, transfers and issuer operations may be delayed or fail. Operational impairment can quickly translate into financial impairment: reduced transferability undermines the stablecoin's usefulness as a medium of exchange, and impaired redemption can widen secondary-market discounts and amplify run dynamics. In addition, issuer-controlled smart contracts introduce issuer-side operational risk, including software defects and weak administrative controls over minting and burning. For example, in October 2025, Paxos mistakenly minted \$300 trillion of PYUSD and corrected the error with an offsetting burn within minutes; Aave paused PYUSD activity and PYUSD briefly traded about 0.5\% below par before stabilizing.\footnote{The erroneous mint of 300 trillion PYUSD happened on the Ethereum blockchain on October 5th, 2025 at 7:12:23 PM UTC and the related burn of 300 trillion PYUSD happened at 7:34:35 PM UTC, a span of just under 23 minutes. Mint transaction hash: 0xc45dd1a77c05d9ae5b2284eea5393ecce2ac8a7e88e973c6ba3fe7a18bf45634 Burn transaction hash: 0xaa532ae7f06cccdbdc226f59b68733ae8594464a98e128365f8170e305c34f4b}
The regulatory relevance is that some drivers of impaired transferability and supply control are within the \si's governance (contract design, key management, multi-party approvals, caps and rate limits, incident response), while others arise externally at the blockchain's consensus and network layers.

\textbf{Quantum threats.} Most major blockchains rely on signature schemes such as ECDSA or EdDSA that are not considered quantum-resistant. A sufficiently capable, error-corrected quantum computer running Shor's algorithm could in principle derive private keys from public keys and authorize fraudulent transfers, enabling theft at scale and undermining confidence in the ledger. Although a cryptographically relevant quantum computer (CRQC) is not known to exist and timelines are uncertain, the issue is preparedness for a low-probability, high-impact transition that could disrupt payment continuity if not executed in an orderly fashion. Post-quantum standards are emerging (including NIST post-quantum cryptography standards),
but public blockchains would still require coordinated protocol upgrades, wallet support, and migration of assets to new cryptographic primitives \citep{NISTPQC,pq-dlt}. 
Note that should a CRQC become practical, it would have a broad impact beyond the public blockchain ecosystem.
Impacted areas would also include the traditional financial system as well as communications, government, military, healthcare, and other critical infrastructure.
For policymakers, the relevant question is whether stablecoin arrangements have credible governance and operational plans to execute such a migration without jeopardizing users' ability to transact and redeem at par, especially given that stablecoin issuers and operators do not control the underlying blockchains.

\textbf{Technical risk framework.} There are several categories of \stc risk that can impact par-value.
A \stc can suffer from an unrestricted change in token supply.
This can render the \si insolvent if liabilities are increased without a corresponding increase in assets.
A \stc can suffer network liveness faults where it lacks the ability to process transactions.
This undermines the usefulness of the \stc as a medium of exchange.
A \stc can also suffer from logic bugs, administrative key compromise and other operational issues.
Some effects can be limited to a small number of transactions while others can have universal impact.
Some of these risks are controllable by the \si while others are imposed externally. 




The following table presents a matrix that covers the probability and potential impact of various technical \stc risks.
We classify risks by probability and potential impact in order to clarify the types of technical failures that could cascade into broader financial impact.
We also indicate which risk is controllable by the \si (I), which are externally imposed (E) and which are shared between the two (I/E).
MITRE's AADAPT knowledge base IDs, which categorize adversary techniques for digital asset management systems, are included if available.\footnote{MITRE AADAPT™ (Adversarial Actions in Digital Asset Payment Technologies) Framework: https://aadapt.mitre.org}

High systemic risk relates to issues which impact both the control of the supply and the liveness of the entire system.
Faults in this class can impact all users and result in the most financial loss.
Low systemic risk only impacts liveness and does not pose uncontrolled supply issues.

\begin{figure}[H]
\small
\begin{tabularx}{\textwidth}{|p{1.5cm}|X|X|X|}
\hline
\cellcolor{gray!25}& \cellcolor{gray!25} \textbf{Most Likely} &\cellcolor{gray!25}\textbf{Moderately Likely} &\cellcolor{gray!25}\textbf{Least Likely} \\
\hline
\textbf{High Systemic Risk}\cellcolor{gray!25}&
\begin{itemize}[nosep,leftmargin=*,topsep=0pt]
  \raggedright
  \item Smart Contract Logic Flaws (I) AADAPT: ADT3012
  \item Bridge Failure (I/E) AADAPT: ADT3029
\end{itemize} &
\begin{itemize}[nosep,leftmargin=*,topsep=0pt]
  \raggedright
  \item Flash loan pricing attacks (E) AADAPT: ADT3015
  \item Administrative key custody failures (I) AADAPT: ADT1552
  \item Contract Upgrade Failures (I)
\end{itemize} &
\begin{itemize}[nosep,leftmargin=*,topsep=0pt]
  \raggedright
  \item Consensus Attacks (E) AADAPT: ADT3013
  \item Cryptographic Exploits (E)
  \item Compiler bugs (I/E)
  \item Integration bugs (I)
  \item Validator coercion (E)
\end{itemize} \\
\hline
\textbf{Medium Systemic Risk}\cellcolor{gray!25}&
\begin{itemize}[nosep,leftmargin=*,topsep=0pt]
  \raggedright
  \item Centralized Infrastructure Failure (I/E) AADAPT: ADT3008
  \item DDoS Attacks (E)
\end{itemize} &
\begin{itemize}[nosep,leftmargin=*,topsep=0pt]
  \raggedright
  \item Governance Attacks (E)
  \item MEV Exploitation (E)
\end{itemize} &
\begin{itemize}[nosep,leftmargin=*,topsep=0pt]
  \raggedright
  \item Replay attacks (I/E)
\end{itemize} \\
\hline
\textbf{Low Systemic Risk}\cellcolor{gray!25}&
\begin{itemize}[nosep,leftmargin=*,topsep=0pt]
  \raggedright
  \item Liquidity Pool Imbalances (E)
  \item Front-Running (E)
  \item Slippage beyond tolerance windows (E)
\end{itemize} &
\begin{itemize}[nosep,leftmargin=*,topsep=0pt]
  \raggedright
  \item L2 Specific Vulnerabilities (I/E)
  \item Gas Price Manipulation (E)
  \item Block stuffing to force higher fees and/or censorship (E)
\end{itemize} &
\begin{itemize}[nosep,leftmargin=*,topsep=0pt]
  \raggedright
  \item Block Reorg Attacks Time-based Vulnerabilities (E) AADAPT: ADT3003
  \item Function signature mismatches (I/E)
\end{itemize} \\
\hline
\end{tabularx}
\caption{Technical Risk Matrix}
\label{fig:risks}
\end{figure}

Smart contract logic flaws can result in uncontrolled supply and transactional liveness issues.
Upgradable \stc smart contracts are also vulnerable to logic issues introduced via the upgrade process.
They are vulnerable to upgrade credentials becoming known externally, resulting in a catastrophic loss of issuer supply control and network liveness.
Both of these risks are within the control of the \si but must be mitigated before contract deployment.

Bridges operate across \stc networks and include the ability to alter the supply on these networks.
Loss of a bridge can cause deposit and redemption liveness issues while bridge takeover or key exposure can result in uncontrolled supply.
Control over this risk exists when the bridge is operated by the \si, but some bridges are operated externally.

A flash loan is a form of riskless lending that is taken out and repaid in the same moment which can be used to amplify an economic attack.
They can be leveraged to increase the magnitude of price dislocations in decentralized markets.
This could be used in extreme cases to alter the value of base layer tokens impacting consensus incentives, or alter the value of reference prices in algorithmic \stc{s}.
Flash loans are unique to blockchains and not within the control of the \si.

Consensus and validator attacks primarily result in censorship and transactional liveness faults and are outside the control of the issuer.
However, in the extremely rare case of a > 2/3 consensus takeover of a PoS chain by an attacker with large compute resources, uncontrolled supply is also possible.
In this case, an attacker obtaining keys from exited validators can create an alternative history including altered mint transactions.
However, this would trigger a detectable chain split which should be rejected by the community.

Smart contract compiler and integration bugs are not very likely but pose both transactional liveness and uncontrolled supply issues, making them highly systemically risky.
However, like smart contract bugs, mitigation of this risk exists with the \si.

Cryptographic exploits and network layer compiler bugs impact both blockchain and traditional systems.
The recourse in a permissionless blockchain based system involves human governance actions which are not clearly defined.
These issues impact both transactional liveness and the control of supply.

Validator coercion impacts transactional liveness and in extreme cases, control over the supply.
Both blockchain based and traditional systems suffer from centralized infrastructure failure and distributed denial of service (DDoS) attacks.
Governance attacks primarily impact supply control and this risk is outside the control of the \si.
MEV exploits are a type of deliberate front-running which also primarily impacts transactional liveness.
Replay attacks where a single transaction is valid on more than one blockchain are not very likely but possible if a contentious hard fork of the network is made without adequate countermeasures.

All of the other low systemic risks are outside the control of the issuer.


\subsection{Technical risk mitigation}

A \si can reduce its controllable risks on public permissionless blockchains through practices including testing, auditing, monitoring, and implementation of operational best practices.
These measures lower the likelihood of failure, but they cannot completely eliminate technical risk.
We now turn to measures that reflect industry best practices for the development, testing, deployment, and monitoring of a \si{'s} smart contracts.

\subsubsection{Smart contract layer and administration}

Stablecoins are implemented using smart contracts, which are software programs executed by the blockchain's nodes.
As \si{s} develop the underlying smart contracts, they must properly architect these programs and mitigate the risk of introducing bugs that could impair the \stc{'s} functioning.
Software bugs are among the most fundamental and significant potential vulnerabilities in a stablecoin's deployment and use.
Typically, \si{s} must first conduct internal testing of the source code, followed by external assessments by independent smart contract auditing firms, before the code can be deployed live to support transactions.

The \si should also maintain distinct environments for development, staging, and production, and consider how they administrate the smart contracts, including key management and access controls.

\textbf{Development and testing.} As developers contribute to the smart contract code base, comprehensive testing should be employed.
This testing of the underlying code must be thorough and evaluate each unit (a single component of the smart contract), its integration into the \si's broader platform (testing of the interaction between units), and fuzz testing (probing how the system behaves when it receives unpredictable inputs).
Formal verification for critical contract components is encouraged where practical.

After internal testing is complete and before deployment or upgrade, \stc smart contracts should be audited by multiple independent smart contract auditing firms to identify any remaining software bugs.
Using common, well-understood libraries reduces auditing demands by shrinking the attack surface, since those libraries have already been extensively audited.

\textbf{Deployment.} It is an industry best practice for the \si to deploy an upgradable smart contract so it may be revised later to address potential bugs or support future expansion.

To avoid the introduction of erroneous or malicious code, upgradable contracts may be deployed with appropriate time locks and multi-signature requirements.
Time locks delay the smart contract's actual deployment, creating a window for additional review, allowing identification of irregularities and for response systems to mitigate them.
Multi-signature arrangements require approval from multiple internal parties to activate the new contract.
By bringing in multiple reviewers from different parts of its operations, the \si can reduce the risk of inadvertently deploying bugs to the blockchain. 

\textbf{Circuit breakers.} Stablecoin issuers need administrative functions such as the ability to halt live operations in the event of significant anomalies.
While issuers build manual kill switches into the smart contract code, they should also develop circuit breakers that automatically stop transactions that are clearly out of bounds, such as minting an exceptionally large number of tokens.
These stopgap functions constrain the magnitude of damages once a process runs awry.
The \si may gradually increase caps and limits as the system proves its stability.

\textbf{Administrative and compliance operations.} Once a smart contract is deployed, the \si must be able to perform administrative tasks, such as minting, burning, or pausing transactions.
They also must be able to manage the blocklist of addresses that are not permitted to be involved in transfers of the \stc, and to upgrade the contract when needed. There is a misunderstanding about what it means when it is said that blockchain-based stablecoin transactions are "peer-to-peer." Note that the \si's smart contracts must process any on-chain stablecoin transaction. Currently, \si{s} are not required to intermediate stablecoin blockchain transactions. This is a point where a \si could intermediate in transactions, if compelled to do so.

\textbf{Administrative access controls.} To carry out the above operations safely, issuers should use robust access controls, such as multi-signature approval schemes with appropriate thresholds and a diverse set of signers who approve changes to those operations.
Time-locks can provide an additional safeguard for sensitive actions, and clear key-management procedures, including the use of hardware security modules (HSMs), help prevent operational errors.
Separating routine operational keys from emergency-only keys further limits the impact of any compromise of internal controls.
Defense-in-depth security practices, regular assessments of administrative infrastructure, and careful documentation of all administrative actions contribute to a more resilient stablecoin system.

\textbf{Bug bounties.} Even after developing, testing, and auditing new code, some problems may remain undetected until the code is deployed on a blockchain.
Any external party can audit smart contracts because they are publicly visible, which increases trust in the system and encourages more reviews of the underlying code.
Nonetheless, any issues public stakeholders identify can be rapidly exploited if the external party that spots a problem does not share this information with the \si first.
Incentives for the responsible disclosure of any problems external parties may discover increases the speed at which the \si can address vulnerabilities in the code.
This can be partially mitigated by maintaining bug bounty programs with meaningful rewards.

All of these strategies reduce the incidence of software bugs, whether introduced erroneously or maliciously.

\textbf{Bridges.} A bridge smart contract enables transfers of a \stc between supported blockchains.
The ability to move \stc{s} from one blockchain to another expands the range of counter-parties with whom a \stc holder can transact.
Some stablecoin issuers run their own cross-chain services to move stablecoins, but there are many third-party bridges which do not appear to have contractual relationships with stablecoin issuers, and to-date, stablecoin issuers do not necessarily license or permission bridges and thus do not have influence into how they operate. 

Bridges function by synchronously swapping the ownership of funds on one chain with an equal amount on another.
They may do this by accepting assets into the control of the bridge operator on one chain and pay that same amount of assets out of their inventory to the same customer on another blockchain.
Alternatively, if operated by the \si they may burn value on one blockchain and recreate that same value on another blockchain.
In these cases where bridges automatically mint and redeem assets on multiple blockchains, they represent a high systemic risk to the \si.

As with smart contract upgrade concerns, a large degree of this bridging risk can be mitigated in the same ways: bridge builders must conduct thorough testing, use third-party auditing, and may employ time locks, but solutions must be automated and avoid routine manual interventions.
For example, if a transaction exceeds a certain threshold, the \si could time-lock the transfer so that it is unusable for a period, allowing for mitigation of a mistaken or malicious attempt to initiate a large transfer of stablecoins.
Additionally, a \si may implement an upper-bound limit on the amount any particular \stc can be bridged over some limited time frame.

\subsubsection{Blockchain layer}
\label{sec:blockchain_layer}


Stablecoins run on public, permissionless blockchains, whose primary feature is that they are not governed by a single or small set of actors.\footnote{Though this is the stated goal, several so-called permissionless blockchains are in fact controlled by a small set of actors; usually a non-profit foundation partnered with a for-profit company. These actors often hold significant amounts of the blockchain's native token, run or strongly influence most of the validators in the network, and thus can direct or coordinate changes to blockchain-native issuance and protocol upgrades. Interestingly, in these cases some risks are exacerbated (it is easier for validators to collude to censor) while others are less of a concern (in the event of a bug or attack, it is easier for validators to cooperate to roll back errant transactions). It is common for these organizations to compensate the largest stablecoin issuers to get them to issue their stablecoin on their blockchains, implying the potential for other types of contractual relationships, perhaps to manage risk.} This is quite different than how the traditional financial system operates, and means that there are many technical risks that are not under a \si{'s} control.

\textbf{Consensus.} A blockchain’s consensus process determines which transactions become part of the authoritative ledger and when they can be treated as final. In proof-of-stake\footnote{Most major stablecoins operate on proof-of-stake blockchains.} systems, validators (independent operators running specialized software) check transactions against protocol rules, assemble them into proposed blocks, and then attest to whether those blocks should be accepted by the network. For stablecoin payments, this layer is not a technical detail: it directly governs settlement finality, throughput, and the conditions under which transfers can be delayed or excluded.

Consensus security depends on the validator set being meaningfully decentralized. Because voting power can be concentrated through stake, a single entity can effectively ``collude'' by controlling many validators, even without coordination among separate firms. If a dominant validator or cartel gains sufficient influence, it can censor otherwise valid transfers, reorder transactions for profit, or weaken reliability during stress---all of which can impair payment continuity and undermine confidence in \stc par-value exchange. We explain this dynamic in more detail in Section~\ref{sec:stablecoins_blockchains}.

Risk mitigation starts with chain selection and operational discipline. \SI{s} should monitor stake and voting concentration, governance changes, and validator behavior that could indicate censorship or instability. They should apply conservative finality requirements (and, where appropriate, longer confirmation thresholds) before crediting high-value redemption transfers. While proof-of-stake protocols can penalize misbehavior through slashing, this is not a substitute for prevention. Resilience is strengthened by diverse validator participation, multiple independent client implementations (to reduce correlated software faults), and clear escalation channels with major validators and ecosystem operators when rapid coordination is required.


\textbf{Monitoring.} Stablecoin operations require continuous monitoring and an incident-response capability that is credible under stress. \SI{s} should treat blockchain-facing operations as critical payment infrastructure. They should maintain real-time visibility into network health (finality delays, congestion, reorg risk signals), contract and key activity (mint/burn events, admin calls), and dependencies such as custodians, wallet integrations, and Remote Procedure Call (RPC) providers. Equally important are predefined response playbooks, secure communications for rapid coordination, and routine security testing (including independent assessments and controlled penetration exercises) to validate that controls work in practice.

Monitoring should also cover consensus-layer attack patterns that may not appear as conventional software failures. For example, ``long-range'' attacks in proof-of-stake designs can arise if keys from past validators are compromised and used to sign an alternative historical chain in which the attacker has no stake at risk. While modern protocols include defenses, early detection of anomalous reorganizations, inconsistent checkpointing, or unexpected finality behavior is essential to limit user harm, preserve operational continuity, and support timely supervisory reporting.


\textbf{Infrastructure.} To issue and administer an on-chain offering, \si{s} must operate infrastructure that can reliably observe the blockchain and submit transactions (routine and administrative). This infrastructure should be redundant, geographically distributed, and designed to fail over safely such that the issuer does not lose the ability to redeem, manage supply, or respond to incidents during localized outages or provider failures.

Most issuer interactions occur through nodes and RPC endpoints, which can become choke points during market stress. Infura, for example, provides widely-used RPC endpoints. Public-facing endpoints should be protected against denial-of-service attacks, and issuers should maintain multiple, independent RPC and node providers (or self-hosted capacity) with regularly tested failover and disaster-recovery procedures. Routine resilience testing matters as much as architecture: backup systems that are not exercised may fail when needed.

Finally, operational controls around privileged access are central to safety and soundness. Administrative keys and infrastructure accounts should be governed by strong access controls, separation of duties, robust logging, and secure custody arrangements (e.g., multi-party approval and hardened key storage). These measures reduce the likelihood that a single compromise, configuration error, or insider action escalates into a supply-control incident or a prolonged disruption of payments---but they do not eliminate risk, underscoring the importance of layered defenses.

\section{Interactions between financial and software risks}
\label{sec:interactions}

In this section we discuss how incentive risks relating to how a blockchain and stablecoins interact could cause additional problems. First, we emphasize that attacks on a blockchain will affect stablecoin management and redemption even if the issuer is not in financial distress; an attacker could censor administrative actions on the stablecoin's smart contract or user transactions trying to redeem, affecting \pv.
Second, we discuss how a stablecoin with large transaction volumes might increase the incentive for an attacker to subvert the underlying blockchain, bringing forth these very attacks.
Third, we point out that large stablecoin issuers, like any centralized custodian of tokenized assets, can have influence on blockchain governance, threatening the prospect of credible neutrality.
Finally, we point out DeFi's dependence on stablecoins and how an attack on a widely used stablecoin might have cascading negative effects on DeFi protocols.

\subsection{Stablecoins' vulnerability to underlying blockchain attacks}
\label{sec:stablecoins_blockchains}

An inability to complete transactions in a timely fashion compromises the usefulness of \stc{s} as a medium of exchange. A loss of liveness, or an increase in the perceived risk of such a loss, can undermine confidence in the \stc and trigger a redemption run. This can overwhelm the ability of the \si to maintain \pv, for reasons discussed in Section \ref{sec:redemption_risk}. Moreover, insofar as a loss of liveness affects all transactions on a blockchain, it can trigger a correlated redemption run among all \stc{s}.
As discussed in Section~\ref{sec:blockchain_layer}, the underlying blockchain consensus protocol determines which transactions are processed and in what order. Therefore, stablecoin management and transfers could be impeded by attacks on the underlying blockchain.
For the purposes of illustration, we focus on blockchain attacks on Ethereum and its underlying PoS protocol, but it is important to note that these types of attacks are not unique to Ethereum. 

There are several classes of attacks on Ethereum's post-Merge PoS protocol which have been studied and identified that would degrade liveness or impede transaction finality. Among them are (i) timing-based and vote-propagation attacks that exploit minor network delays to prevent the formation of a supermajority and a balancing attack and that creates uncertainty about which of two competing chains will ultimately finalize by eclipsing a portion of the network \citep{neu2020balancing,schwarzschilling2021three}, (ii) secret-chain reorganization attempts, in which an attacker with a supermajority of stake privately constructs an alternative chain and releases it strategically when the financial incentive is greatest \citep{neuder2021subthird} and (iii) public competing-chain attacks, in which an attacker broadcasts an alternative chain in real time to generate temporary uncertainty over finality \citep{schwarzschilling2021three}. 

There is a particularly pernicious attack that is specific to \stc{s} and which has not (to our knowledge) been heretofore studied. It involves a combination of a  double-spend attack combined with censoring an administrative transaction from a \si to suspend the transfer by blacklisting the attacker's address or pausing the overall contract. If an attacker can censor the \si administrative action, it can act unopposed. The goal of the attacker in each type of attack is to induce the victim to deliver her item of exchange to the attacker and then to nullify the transfer of ETH or the stablecoin. Notably, in some cases the attacker can accumulate transactions before triggering the nullification, with the attacker double-spending several counterparties. 

An attacker could censor transactions or halt the chain (affecting liveness) with only slightly more than 1/3 of the underlying stake. An inability to complete transactions in a timely fashion compromises the usefulness of \stc{s} as a medium of exchange. A loss of liveness, or an increase in the perceived risk of such a loss, can undermine confidence in the \stc and trigger a redemption run. This can overwhelm the ability of the \si to maintain \pv, for reasons discussed in Section \ref{sec:redemption_risk}. Moreover, insofar as a loss of liveness affects all transactions on a blockchain, it can trigger a correlated redemption run among all \stc{s}. 

Table \ref{tab:attacks} in Appendix~\ref{app:attacks} compiles a pattern of significant declines in cryptocurrency prices following attacks on the associated blockchain (the examples are not limited to Ethereum). This response reflects a loss of confidence in the integrity of the blockchains.  It is plausible that a \stc operating on an attacked blockchain would experience a similar loss of confidence from an attack. 

A disruption to the functioning of the blockchain will disrupt the execution of \stc transactions that are dependent on the functioning of the blockchain. An attack severe enough to cause a loss of confidence in the underlying blockchain token might cause a loss of confidence in the \stc by at least some percentage of holders, who would exit by redeeming their \stc{s}. 


\subsection{An increase in stablecoin volume can incentivize attacks on the blockchain}



The attacks described above rely on a large amount of stake colluding to conduct the attack. The incentive to collude increases as the potential reward increases. This reward is the amount gained by disrupting the blockchain (a double spend or the profits from a short on ETH) minus the cost to conduct the attack (primarily, the cost of the slashed stake). Note that the latter is denominated in the underlying blockchain token, ETH, while the former could be denominated in the stablecoin. As stablecoin transaction volume rises, the opportunity for double spends in the stablecoin increases, while there is not necessarily a corresponding increase in the punishment for conducting an attack. The cost to mount the attack does not scale with the economic value transacted on Ethereum, but the expected reward does. 

The result is a structural vulnerability: as stablecoin activity scales, the prospective gains from manipulating transaction settlement, delaying finality, or creating transient forks grow more rapidly than the honest-validator reward rate. Even if the likelihood of these attacks remains low under normal conditions, the economic incentive gradient steepens with stablecoin adoption. The combination of attack vectors that degrade liveness or obscure finality and the rising value at stake in stablecoin transactions implies that Ethereum’s post-Merge PoS chain will become increasingly attractive to adversaries as stablecoins scale.

\subsection{Large stablecoin issuers might influence blockchain governance}


As others have noted, \si{s} might have unexpected influence on blockchain upgrade and governance decisions because in the event of a blockchain fork due to disagreement, they can choose which side of the fork to honor for stablecoin redemptions~\citep{lee_qureshi_ethereum_unforkable_2019, narula_stablecoins_2021}. It is unclear how this might influence governance decisions as stablecoins grow ever larger, and potentially even eclipse the monetary base of the underlying blockchain's native token. If a large \si has this influence, it reduces the credible neutrality of the underlying blockchain as an infrastructure layer. The \si might be required to choose the fork in alignment with the legal jurisdiction in which they operate. 



\subsection{DeFi's dependence on stablecoins}



Stablecoins are widely used in DeFi smart contracts. For example, there is over \$270 million of USDC in Uniswap pools on Ethereum~\citep{UniswapUSDC}.
If USDC were to suffer an attack, there might be a run on the stablecoin and many people might rush to withdraw from these pools. Other work has shown that this can serve as a channel for financial contagion in the DeFi sector~\citep{Du-shadowbankruns2025}.

\subsection{Summary}
Financial and software risks reinforce one another in ways that can undermine the stability of a \stc. Disruptions to the blockchain can delay redemptions or transfers, weakening confidence in the \stc’s reliability as a medium of exchange and amplifying run risk. Rising \stc volume increases incentives to attack the host blockchain by raising the prospective gains from controlling transaction processing. Large stablecoins can influence blockchain governance and cause vulnerabilities for DeFi protocols. Together, these channels illustrate how blockchain-level vulnerabilities can endanger the financial stability of \stc{s}, and vice versa. Crucially, many of the vulnerabilities reviewed here are outside the control of the \si and therefore cannot be mitigated by the \si. They arise from the consensus mechanism that underlies a permissionless blockchain.


\section{Regulatory safeguards and outstanding policy questions under GENIUS}
\label{sec:safeguards_questions}

\subsection{Strengths of the GENIUS framework}

As addressed earlier in this paper, the GENIUS Act establishes the initial federal architecture for issuing and supervising payment stablecoins. Its strengths lie primarily in its efforts to shape the financial infrastructure underpinning a stablecoin. The Act defines licensing requirements for payment stablecoin issuers, imposes a strict reserve-backing asset requirement, and identifies eligible instruments for reserve backing assets. These constraints sharply limit credit risk and restrict exposure to long-duration securities, which, in turn, may enhance solvency and liquidity, respectively. Monthly public disclosures of reserve backing asset composition and valuations further strengthen transparency and allow market participants to assess issuer risk on an ongoing basis. GENIUS's prohibition against the rehypothecation of customers' assets and its requirements for independent audit attestations reduce the risks that opaque leverage or undisclosed exposures build up inside backing asset portfolios.

While the GENIUS Act is now law, regulatory agencies must transpose its principles into tangible regulation, and several provisions in the Act explicitly require further study. Relevant U.S. regulatory agencies must interpret the statute, issue regulations, and build a framework for licensing, supervision, enforcement, and technical standards.  Beyond immediate prudential rules, the Act mandates several formal studies, including analyses of potential capital requirements, oversight structures, and the financial stability implications of stablecoin growth. These provisions acknowledge that refinements may be necessary to the regulatory framework as stablecoin adoption expands and as the interaction between market infrastructure and digital asset regulation becomes clearer.

\subsection{Remaining gaps, structural dilemmas, and unresolved questions}

While GENIUS addresses reserve backing asset quality and transparency, it leaves several foundational issues unresolved. 

First, although the GENIUS Act requires issuers to disclose a redemption policy and to establish procedures for the timely redemption of stablecoins \citep{12USC5903}, it does not specify how redemptions should work. As discussed in Section~\ref{sec:key_elements}, this is problematic because most stablecoin holders today cannot redeem their coins directly with issuers. Instead, holders rely on intermediaries such as cryptocurrency exchanges (e.g., Binance or Coinbase), where par value redemption is not assured. At market exchanges, prices for redeeming stablecoins fluctuate with supply and demand. 

Some issuers do offer 1:1 redemption to holders who have verified financial accounts with the issuer, but that access is tightly restricted. For example, Circle limits direct redemption to 521 entities and Tether limits redemption to six entities \citep{ma2025stablecoin}. These entities act as minting and redemption intermediaries between the \si and other owners of \stc{s}. 


If regulators permit the current approach to redemptions to continue, the existing two-tiered system will remain. Institutional clients (such as those that have verified accounts with stablecoin issuers) will be able to redeem their coins directly with the issuer, most likely at par. Others, such as most retail customers, will depend on the peg holding in the secondary market. In times of market stress, the value of a stablecoin may fall below par in secondary markets, as it did for USDC during the Silicon Valley Bank crisis \citep{huang2023circle}. The only way out for many retail owners of stablecoins in such situations may be to sell their coins at a loss on the open market.

Moreover, frictions arise when issuers must liquidate U.S. Treasury securities through broker-dealer intermediaries that are constrained by balance sheet and leverage ratio requirements. These constraints can impede orderly liquidation precisely when redemptions surge.

In light of recent research documenting systematic and persistent deviations from par in secondary markets \citep{ma2025stablecoin}, the GENIUS Act's omission of any reference to maintaining par-value exchange in the secondary market (where most stablecoin users actually transact) raises concerns about whether stablecoins will maintain the public's confidence over the long run. 

Second, although the Act requires further study of capital standards, it appears to restrict U.S. regulators from applying directly to stablecoin issuers the minimum leverage and risk-weighted capital requirements that govern commercial banks, as noted in Section~\ref{sec: Balance Sheet Resilience}. Likewise, GENIUS is silent on the dilemma analyzed in Section \ref{subsec:dilemmas}: whether stablecoin issuers should have access to Federal Reserve liquidity facilities. Granting such access could materially reduce redemption-related liquidity bottlenecks. Yet doing so raises complications for monetary policy transmission, the boundary between banks and nonbanks, and the distribution of reserves in the financial system.

Even though stablecoin issuers face run-risk dynamics akin to banks and money-market funds, the GENIUS Act potentially enables a regulatory asymmetry: stablecoin issuers are expected to maintain \pv but are not required to adopt structural protections such as capital buffers or be permitted to access public backstops that support issuers of other forms of demandable liabilities that are expected to honor \pv exchange.

Third, the GENIUS Act authorizes regulators to prescribe interoperability standards for stablecoins, but it does not define what interoperability actually means in practice. As a starting point, interoperability can be broken down into two components, namely network interoperability versus monetary interchangeability.

\textbf{Network Interoperability} asks whether a stablecoin can be used across multiple blockchains, wallets, and platforms without friction. Because the GENIUS Act does not require issuers to support technical standards across chains or wallets, the Act's silence on the matter creates a risk that stablecoins will operate in isolated silos. For example, a stablecoin on Ethereum might not interoperate with the Solana version of that stablecoin. 
The BIS has warned that a stablecoin issued on different blockchains may not interoperate cleanly \citep{cpmi2023stablecoin}. Without guidance, the U.S. could end up with a digital dollar landscape badly fractured by incompatible technical standards, which could affect liquidity depth and user experience.

\textbf{Monetary Interchangeability} asks whether different USD-based stablecoins are exchangeable with each other at par. Beyond technical standards, however, lies a deeper issue of monetary interchangeability. This is the idea that all stablecoins, like all dollars, should be usable at equal value: ideally a dollar should always be worth a dollar, regardless of its form, whether it is held as a bill, in a bank account, or as a stablecoin. There are many structures in our monetary system to try to ensure this for commercial bank deposits (though it doesn’t always hold all the time). The BIS argues that many stablecoins fail this “singleness of money” test because they are not always redeemable at par or interchangeable across users~\citep{Shin2025BIS}.

The GENIUS Act does not address

\begin{itemize}
    \item whether stablecoins from different issuers should be fungible, or
    \item how to ensure \pv exchangeability between similar stablecoins. 
\end{itemize}

This omission opens the door to a potentially fragmented and fragile payment landscape. Some observers compare this potential outcome to the ``Wildcat Banking'' era in U.S. history, when private banks issued their own banknotes that traded at varying discounts \citep{gorton2023taming}. A more nuanced view by \citet{carter2025lastword} suggests stablecoins could evolve toward a model of competitive digital dollars, where interoperability standards and issuers determine trust. Such an outcome might deviate from the singleness of money in two respects: First, there would be no assurance of acceptance of any given stablecoin as payment, and second, differences in valuations would imply deviations from par-value.

Finally, the GENIUS Act is silent on the technical infrastructure underpinning stablecoins. To date, stablecoin issuers decide for themselves whether to obtain their own technical audits and may claim to follow best practices, but this approach lacks transparency, accountability, and enforcement. The GENIUS Act advances no requirements for technical audits, open-source transparency, or even secure key management. Furthermore, the stablecoin ecosystem is highly integrated across many infrastructure layers, like wallets and bridges, which have been frequent targets of attacks in crypto markets. Without minimum risk standards for these technical integrations, GENIUS Act-compliant stablecoins may still be vulnerable. It remains to be seen whether regulatory agencies will fill those gaps to strengthen public confidence in stablecoin issuers and the underlying infrastructure they depend upon.

Failing to address these technical concerns risks creating a stablecoin that may be legally compliant, but technically insufficient. Where something does go wrong, there’s no clarity about who might be responsible or whether redemptions would still be enforceable. 

In summary, GENIUS establishes a strong foundation for backing asset safety and transparency. It leaves open many critical questions about liquidity support, redemption mechanics, capital adequacy, interoperability, technical infrastructure, and the long-run integration of stablecoins into the monetary system. In fairness, regulators have not yet issued proposed rulemakings at the time of this writing, and, consequently, they may mitigate some of the challenges we note here. This discussion above serves as an (incomplete) list of additional areas that need to be addressed.


\section{Conclusion}
\label{sec:conclusion}

Stablecoins promise to expand the reach of dollar-denominated payments by combining the monetary reliability of traditional assets with the programmability and interoperability of digital networks. The GENIUS Act establishes the first comprehensive federal framework governing these instruments, setting clear conditions for issuance, backing assets, and transparency. Yet the analysis in this paper shows that the stability of GENIUS-compliant stablecoins hinges on how financial risks, technological risks, and regulatory constraints interact—especially under stress.

We first demonstrated that the core financial vulnerability lies in the tension between \pv redemption and market intermediated liquidation of Treasuries. Stablecoin issuers can mitigate the risk of capital loss by investing in reverse-repos. However, to the extent that \si{s} hold Treasury securities, their ability to redeem at par depends on executing timely asset sales through broker-dealers whose balance sheets are themselves constrained by leverage rules and liquidity regulations. Episodes of fragility in the Treasury and repo markets show that even modest surges in selling pressure can overwhelm intermediation capacity due to the fact that the bank affiliates of the largest broker-dealers operate near the regulatory lower bound of their supplementary leverage ratio. This limits their ability to increase balance-sheet assets to accommodate purchases of Treasuries, either by creating a \bd for the \si, or by obtaining central bank reserves, from the Fed's standing repo facility,  to transmit to the \si{'s} bank. Each action pushes down the supplementary leverage ratio. These structural bottlenecks create the possibility of price deviations, delayed redemptions, or runs—especially when redemption demand rises rapidly.

We next extended this analysis by showing that the \textit{technological substrate} of stablecoins introduces a qualitatively different layer of risk. Smart-contract design, bridge and oracle architectures, validator incentives, governance structures, and the liveness of public blockchains all shape a \stc{s} operational reliability. Failures in these components—whether from logic bugs, key mismanagement, consensus faults, or adversarial attacks—can impair users’ ability to transfer or redeem tokens even when backing assets remain intact. Moreover, the value transacted in stablecoins can itself amplify incentives to attack the underlying blockchain, creating a two-way linkage between financial scale and cyber vulnerability. These mechanisms illustrate that operational failures can propagate into financial instability through loss of confidence, impaired liquidity, or sudden redemption pressures.

Finally, we evaluated how the GENIUS regulatory framework addresses these risks---and where it does not. The Act succeeds in strengthening asset quality, transparency, and supervisory oversight. Its prohibition on rehypothecation of backing assets and its monthly public disclosures meaningfully reduce hidden leverage and opacity. Yet, key dilemmas remain unresolved. GENIUS appears to forbid explicitly the direct application of minimum bank-style capital requirements to issuers, even though they issue demandable liabilities vulnerable to run dynamics. It also leaves redemption mechanics undefined and does not address the liquidity bottlenecks created by dealer balance-sheet constraints. Most importantly, it does not resolve the policy dilemma outlined in Section 5.4: whether stablecoin issuers should have access to Federal Reserve liquidity facilities. Such access could materially improve resilience during stress but carries implications for the implementation of monetary policy and the banking system.

Taken together, our findings show that achieving durable stability in the stablecoin sector will require more than rules that delimit allowable backing assets. It will require an integrated approach that accounts for financial-market structure and blockchain-level risks. It requires careful consideration of the difference between a bank and a nonbank issuer of money-like claims. GENIUS provides a strong foundation, but further policy development—guided by the studies it mandates—will be necessary to ensure that stablecoins can function reliably as scalable, dollar-denominated payment instruments in both normal and stressed conditions.


\section{Acknowledgements}

We are grateful for the excellent research and analytical support provided by Jenny Jin and thank the following individuals for their insight, suggestions, and feedback: Pablo Azar, Robert Bench, Pavel Chichkanov, Eziechel Copic, Darrell Duffie, Ethan Heilman, Nellie Liang, Tim Massad, Susan McLaughlin, and Madars Virza. 

This work was supported by funding from the MITRE Corporation. The content, analyses, and conclusions presented in this report are those of the authors and do not necessarily reflect the views, positions, or policies of the MITRE Corporation or any affiliated organization.


\appendix
\section{Appendix: Leverage Ratio Calculations}
\label{app:leverage-ratio-calculations}

In Section 4, we analyzed the capitalization of select stablecoin issuers using a capital measure applied by U.S. bank regulators: the leverage ratio. We found sufficient data in public filings from five prominent stablecoin issuers to calculate a simplified version of the most fundamental capital requirement for commercial banks, the minimum leverage ratio, for periods ranging from three to ten recent quarters. 

To calculate the leverage ratio, we applied the requirements set out in the Federal Financial Institutions Examination Council's ``Instructions for Preparation of Consolidated Reports of Condition and Income, FFIEC 031 and FFIEC 041'' as of March 2025, Section RC-R~\citep{FFIEC031_041}.
%
We made assumptions about the stablecoin issuers' data to fit that data into the Basel III leverage ratio standards. For example, we assumed that the U.S. Treasuries and the securities that the five stablecoin issuers reporting purchasing from a third party under an agreement to resell (``reverse repos'') would qualify under the GENIUS Act as acceptable reserve assets, even though regulators have not yet defined the requirements. 

We moreover made simplifying assumptions about the form of the stablecoin issuers' capital. As defined by the Basel III and the FFIEC filing instructions, Tier 1 capital has two components: Common Equity Tier 1 (CET1) and Additional Tier 1 capital (AT1). This latter category consists of common shares and stock surplus, retained earnings, other comprehensive income, qualifying minority interest and regulatory adjustments. Given the unspecified structure of stablecoin issuers' capital, we assumed that all equity (total value of reserve assets - total value of coins in circulation) that each stablecoin issuer reported consists solely of Tier 1 capital. We furthermore assumed that the issuer's total assets are equal to the total value of all reserve backing assets. 

With those assumptions in hand, we then calculated a simplified version of the Basel leverage ratio using the following formula: 

$$\text{Leverage Ratio} = \frac{\text{Total Value of the Reserve Assets} - \text{Total Value of the Coins in Circulation}}{\text{Total Value of the Reserve Assets}}$$

Higher ratios indicate that more capital is available to support a bank's unweighted assets, thereby strengthening the bank's ability to absorb unexpected losses from any source and improving its resilience. Under the Federal Deposit Insurance Corporation Improvement Act of 1991 (FDICIA), U.S. commercial banks must have leverage ratios of at least 4\% to meet minimum requirements to be considered adequately capitalized and at least 5\% to be considered “well-capitalized.” (This latter and most favorable designation was meant to indicate that the risk of failure is low.) 

Commercial banks with leverage ratios below 4\% are considered undercapitalized. This unfavorable designation deteriorates to “significantly undercapitalized” below 3\% and “critically undercapitalized” at less than 2\%. Under FDICIA's “prompt corrective action” rules, U.S. bank regulators typically must close critically undercapitalized banks and put them into receivership within 90 days, setting clear expectations for regulatory intervention \citep{fdic2022enforcement, gao2024bank}. 
Stablecoin issuers lack such safeguards at present.

Three of the issuers we studied, namely Circle, PayPal, and Paxos, reported reserve assets that consist primarily of cash, deposits, and U.S. Treasury-related securities, including reverse repos; as such, these three issuers appear to conform to the limitations set by the GENIUS Act on reserve assets that regulated payment stablecoin issuers are permitted to hold. We cannot confirm that these reserve assets qualify, especially since regulators have not yet defined these steps. The other two issuers, namely Tether and Ripple, held assets such as corporate bonds or even digital assets such as Bitcoin; as such, these two issuers would not meet the requirements set out in the GENIUS Act. 

As indicated in Table~\ref{tab:leverage-ratio}, none of the three firms whose reserve assets appear to conform to the GENIUS Act's requirements would have been considered adequately capitalized under the leverage ratio requirement for commercial banks over the entire period we studied. Two of the three would have been considered critically undercapitalized in all quarters for which they provided data, and the third would have declined from significantly undercapitalized to critically undercapitalized in the most recent two quarters. 

In contrast, Tether met the minimum 4\% leverage requirement in five quarters and would have been considered "well capitalized" (above 5\%) in two of those five quarters. The remaining issuer, Ripple, met the minimum requirement in two quarters, of which one quarter it was well-capitalized using this one measure of capital. 

However, since these leverage ratios predate the GENIUS Act, these ratios reflect the inclusion of some higher-risk assets that historically have not maintained stable value, such as corporate bonds, and volatile digital assets, such as Bitcoin, included in Tether's reserve assets. In fact, as of the second quarter of 2025, Bitcoin was equivalent to nearly 5.5\% of its total assets; if the value of Bitcoin were to drop to zero, Tether's liabilities would exceed its assets, rendering it insolvent on paper. 
As mentioned earlier, Tether has announced plans to introduce a new U.S. dollar-based digital asset, USAT, that will comply with the GENIUS Act's limitations on the kinds of permissible reserve assets it may hold, but this coin has not been issued at the time of this writing. ~\citep{TetherUSAT}.



\section{Attacks on blockchains}
\label{app:attacks}

\begin{table}[H]
\centering
\small
\setlength{\tabcolsep}{4pt}
\renewcommand{\arraystretch}{1.15}

\begin{threeparttable}
\caption{Major Blockchain Attacks: Technical Cause, Value, and Price Reaction\newline{\footnotesize Sources: \protect\citep{buterinDAOvulnerability2016,secDAOreport2017,roninValidators2022,certikWormhole2022,bnbChainEcosystemUpdate2022,coinbaseETCReorg2019}.}}
\label{tab:attacks}

\begin{tabularx}{\textwidth}{@{}
  p{0.20\textwidth}
  p{0.18\textwidth}
  >{\raggedright\arraybackslash}X
  >{\raggedright\arraybackslash}X
  >{\raggedright\arraybackslash}X
@{}}
\toprule
\textbf{Incident (year)} &
\textbf{Main chain / token(s)} &
\textbf{Nature of attack} &
\textbf{Est.\ value stolen / created (USD, at time)} &
\textbf{Approx.\ immediate price response} \\
\midrule

The DAO hack ( Buterin \cite{buterinDAOvulnerability2016}, SEC \cite{secDAOreport2017}) &
Ethereum / ETH &
Smart-contract bug (reentrancy / recursive calls) in The DAO contract, allowing repeated withdrawals before balances updated. &
About 3.6 million ETH; \$50--\$60M at the time. &
ETH \(\$21.50 \rightarrow \$15\) within hours (\(\sim 30\%\) intraday). \\

Ronin bridge hack (Ronin Network \cite{roninValidators2022}) &
Ronin / RON; Axie Infinity / AXS &
Cross-chain bridge validator key compromise; attacker controlled 5/9 validators and forged withdrawals. &
173{,}600 ETH + 25.5M USDC (\(\sim\$540\)--\(\$620\)M). &
RON down \(\sim 20\)--\(22\%\) in 24h; AXS \(\$70 \rightarrow \$64.3\) (\(\sim 8\%\)). \\

Wormhole bridge exploit (Certik blog \cite{certikWormhole2022}) &
Solana / SOL (Wormhole bridge) &
Bridge smart-contract vulnerability on Solana; attacker bypassed verification checks and minted 120k wETH. &
About 120{,}000 wETH (\(\sim\$320\)--\(\$326\)M). &
SOL \(\$111 \rightarrow \$96\) in 24h (\(\sim 13.5\%\)). \\

BNB Chain bridge hack (BNB \cite{bnbChainEcosystemUpdate2022}) &
BNB Chain / BNB &
Cross-chain bridge exploit on BSC Token Hub; forged proof/messages enabled unauthorized withdrawal of \(\sim\)2{,}000{,}000 BNB. &
\(\sim\$570\)M (only \(\sim\$100\)M bridged out). &
BNB down \(\sim 3.5\%\) over 24h, to \(\sim\$281\) after disclosure. \\

Ethereum Classic 51\% attack (Coinbase \cite{coinbaseETCReorg2019}) &
Ethereum Classic / ETC &
51\% / majority-hashpower attack; deep reorgs and double-spends of exchange deposits. &
About 219{,}500 ETC double-spent (\(\sim\$1.1\)M). &
ETC from \(\sim\$5.51\) (24h high) to \(\sim\$4.93\) (low); \(>7.5\%\) daily loss, \(>10\%\) from high. \\

\bottomrule
\end{tabularx}

\begin{tablenotes}[flushleft]
\footnotesize
\item Notes: Figures are approximate and refer to conditions at the time of each incident. “Immediate price response” is the first intraday / 24-hour reaction after public disclosure.
\end{tablenotes}
\end{threeparttable}
\end{table}


\end{document}